\documentclass[%
 reprint, twocolumn, 
 amsmath,amssymb,
 aps, physrev,
floatfix
]{revtex4-2}

\usepackage{graphicx}
\usepackage{dcolumn}
\usepackage{bm}
\usepackage[colorlinks = true]{hyperref}

\usepackage{xcolor}
\usepackage{physics}
\usepackage{lipsum}

\newlength{\ketketwidth}
\newlength{\ketwidth}

\newcommand{\kettstylesep}[3]{
    \settowidth{\ketwidth}{$#2\left|#1\right\rangle$}
    \settowidth{\ketketwidth}{$#2\left.\left|#1\right\rangle\right\rangle$}
    \left|#1\right\rangle#3\hspace{\ketwidth}\hspace{-\ketketwidth}
}
\newcommand{\kett}[1]{
    \left.\mathchoice
        {\kettstylesep{#1}{\displaystyle}{\hspace{0.3em}}}
        {\kettstylesep{#1}{\textstyle}{\hspace{0.3em}}}
        {\kettstylesep{#1}{\scriptstyle}{\hspace{0.3em}}}
        {\kettstylesep{#1}{\scriptscriptstyle}{\hspace{0.25em}}}
    \right\rangle
}

\newcommand{\bbrakettstylesep}[4]{
    \settowidth{\ketwidth}{$#3\left\langle#1\middle|#2\right\rangle$}
    \settowidth{\ketketwidth}{$#3\left\langle\left\langle#1\middle|#2\right\rangle\right.$}
    #4\hspace{\ketwidth}\hspace{-\ketketwidth}\left\langle#1\middle|#2\right\rangle#4\hspace{\ketwidth}\hspace{-\ketketwidth}
}
\newcommand{\bbrakett}[2]{
    \left\langle\mathchoice
        {\bbrakettstylesep{#1}{#2}{\displaystyle}{\hspace{0.3em}}}
        {\bbrakettstylesep{#1}{#2}{\textstyle}{\hspace{0.3em}}}
        {\bbrakettstylesep{#1}{#2}{\scriptstyle}{\hspace{0.3em}}}
        {\bbrakettstylesep{#1}{#2}{\scriptscriptstyle}{\hspace{0.25em}}}
    \right\rangle
}

\begin{document}
		
\title{Liouvillian exceptional points in the emergence of quantum synchronization}
\author{Carlos Ortega-Taberner}
\author{Rosalind E. Cotton}
\author{Paul R. Eastham}
\affiliation{School of Physics, Trinity College Dublin, Dublin 2, Ireland}
\affiliation{Trinity Quantum Alliance, Unit 16, Trinity Technology and Enterprise Centre, Pearse Street, Dublin 2, Ireland}

\date{\today}
\begin{abstract}
  The coupling between self-sustained oscillators, such as lasers or polariton condensates, leads to different steady-states in different regimes. Weak coupling allows the oscillators to behave independently, with different frequencies and uncorrelated phases, while stronger coupling can produce synchronization. In classical or semiclassical theories transitions between these steady-states correspond to changes in the topology of the phase space attractors and occur at generalized exceptional points. We investigate the corresponding changes in a quantum theory of coupled lasers or condensates. We show how the different steady-state regimes of the classical limit give rise to distinct forms for the spectra and eigenmatrices of the slow modes of the Liouvillian. By following the spectral flow between the different regimes we show that the synchronization transition is controlled by cascades of exceptional points in these slow modes, protected by a generalized PT symmetry. Our results show how singularities of the quantum Liouvillian control the different dynamical regimes, and give rise to experimental signatures of the synchronization transition which remain well-defined in the few-particle regime dominated by quantum effects.
\end{abstract}

\maketitle

\section{Introduction}

A limit cycle is a closed isolated trajectory in the phase space of a dynamical system. It describes the spontaneously oscillating steady-state of a nonequilibrium oscillator, such as a pendulum clock or a laser, which appears when the energy input exceeds the damping~\cite{pikovsky2001,balanov2009}. This fixes only the amplitude of the spontaneous oscillations, leaving a free phase which enables oscillators to respond to one another. Owing to this, groups of two or more oscillators exhibit different synchronized and desynchronized states, associated with the presence or absence of common frequencies and phases, as well as nonequilibrium phase transitions between the different states~\cite{pikovsky2001,balanov2009,moroney2021,moroney2023}. 
Such effects have been studied in varied settings ranging from fireflies to laser arrays~\cite{vladimirov2003,pfluger2023}, polariton condensates~\cite{ohadi2018,chestnov2019,wouters2008,eastham2008}, and Josephson junctions~\cite{wiesenfeld1996}.

Exceptional points are well known singularities of linear systems~\cite{heiss2012}, such as oscillating piano strings~\cite{weinreich1977}, where they separate over and underdamped regimes. At the critical damping the normal modes collapse and no longer form a complete basis. The concept of exceptional points can be generalized to classical nonlinear systems where the non-local structure of phase space changes, for example, where a single limit cycle becomes a pair. These generalized exceptional points are singularities of the classical time-evolution operator, and lead to non-reciprocal dynamics, destruction of isochrons, and enhanced and non-Gaussian noise~\cite{weis2025}.

In this work, we investigate the analogous transitions in a quantum system. We consider a realistic quantum model of coupled polariton condensates~\cite{whittaker2009} or lasers~\cite{kreinberg2019,scully1997}, in which there are both single frequency and multiple frequency steady-states. Transitions between these regimes can occur through distinct physical mechanisms, depending on the details of the model and form of the coupling~\cite{ding2019}. For the model we consider dissipative couplings produce a synchronization transition due to the mutual influence between phases of weakly-perturbed limit cycles~\cite{vladimirov2003,pikovsky2001}, whereas reactive couplings produce transitions due to mode competition, i.e., the suppression of gain for one coupled mode by the population of the other~\cite{siegman1986,eastham2008}.

We look for the quantum analogs of the generalized exceptional points found in the classical dynamics, without making semiclassical approximations, by computing the spectrum and eigenmodes of the Liouvillian~\cite{minganti2018a}. We build on recent works considering a single self-sustained oscillator as it approaches the classical limit~\cite{cabot2024,dutta2025}, where the limit cycle~\cite{alaeian2021} appears a parabolic branch of slowly decaying eigenmodes of the Liouvillian. For coupled self-sustained oscillators we find that the different steady-state regimes of the classical limit give rise to different forms for this part of the spectrum, and therefore qualitatively different long-time dynamics, as well as different forms of phase-space distributions in the eigenmodes. By following the spectral flow we show that the transitions between different dynamical regimes are controlled by exceptional points in the slow modes, protected by a generalized PT symmetry. Such exceptional points occur also in the semiclassical theory, but we find significant differences in the quantum regime. For example, while in a semiclassical theory the exceptional points appear at the same coupling strength in all orders of coherence, this is not the case in the quantum theory. Our results show how singularities of quantum dynamics are related to synchronization, and give rise to experimental signatures of the transitions which remain well-defined in the few-particle regime dominated by quantum effects.

There has been great interest in generalizing the concepts of synchronization to quantum systems~\cite{schmolke2026,solanki2026}. The quantum theory of the laser provides an early example showing the nature and emergence of a limit cycle in a quantum system, as explored more generally in \cite{lee2013,benarosh2021,chia2020,dutta2025}. The effects of external driving and coupling between oscillators have been studied in many models \cite{mari2013,lee2013,walter2014,walter2015,chia2020,benarosh2021,koppenhofer2019,jaseem2020,manju2023,hong2024,kehrer2025,lai2025,nadolny2026,roulet2018a,roulet2018b,parra-lopez2020,cabot2021}. Synchronization of quantum oscillators has been treated by analyzing phase correlations, among other measures, in steady-states~\cite{walter2014,walter2015,lee2013}, as well as through dynamical observables. While some works have explored quantum versions of the canonical models of classical self-sustained oscillators, such as the van der Pol and Rayleigh oscillators~\cite{lee2013,walter2014,walter2015,chia2020,benarosh2021}, others consider synchronization for few-level systems which lack classical analogs \cite{cabot2021,zhang2023,parra-lopez2020,zhou2023,koppenhofer2019,roulet2018a,roulet2018b,jaseem2020}. Other important issues include the role of non-reciprocal couplings~\cite{belyansky2025} and extensions to many-body systems~\cite{schmolke2026} such as the Dicke (Tavis-Cummings) model~\cite{eastham2003}. However, it has proved difficult to identify clear diagnostics of synchronization.  One difficulty is that in small systems quantum fluctuations, and indeed noise more generally, produce smoothing of the singular behavior of observables that is present in the classical or noiseless limit~\cite{pikovsky2001,stratonovich1967,moroney2023}. Another is related to distinguishing synchronization from the physics of linear oscillators or other nonlinear effects~\cite{pikovsky2001}. Our work addresses both difficulties by relating the different steady-states which occur in the semiclassical limit to the topology of the slow parts of the Liouvillian spectrum, and the transitions with changes in that topology at generalized exceptional points.  

The remainder of this paper is structured as follows. We begin by introducing the single laser model in Sec.~\ref{sec:single_laser}, discuss the signatures of the limit cycle in the Liouvillian spectrum~\cite{dutta2025}, and connect them to the observable coherence functions. In Sec.~\ref{sec:coupled_lasers} we extend this approach to coupled lasers. Section~\ref{sec:dissipative} considers the case of a purely dissipative coupling, which we argue leads to a synchronization transition, in the sense of that for coupled phase oscillators, while Sec.~\ref{sec:reactive} considers the case of a purely reactive coupling, which leads to mode competition and bistability. In Sec.~\ref{sec:spectral_flow} we consider the general case with both couplings, and show how the spectral flow between the different regimes requires the presence of generalized exceptional points. We connect them to those which occur in the semiclassical approximation in Sec.~\ref{sec:comb-coupl-semicl}. Finally, Sec.~\ref{sec:conclusion} discusses our conclusions.

\section{Single laser and quantum limit cycles}\label{sec:single_laser}

\begin{figure*}[t] 
    \centering
\includegraphics[width=\textwidth]{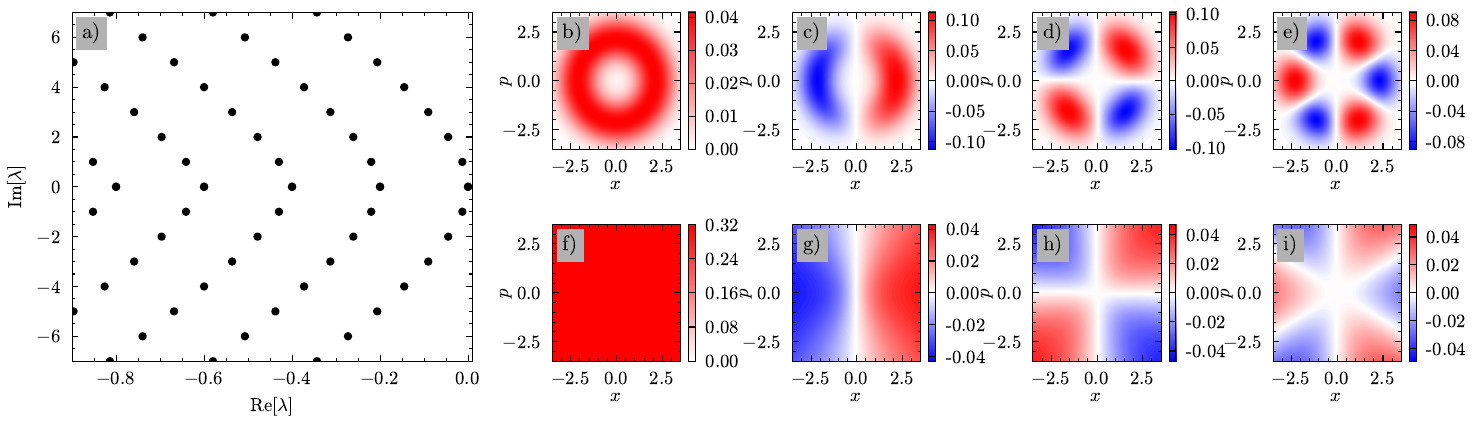}
    \caption{(a) Spectrum of the Liouvillian of the laser, showing different parabolic branches of eigenvalues approaching $\Re[\lambda] = 0$, which characterize the limit cycle. (b-e) Real part of the Husimi function of the least damped right eigenmatrices of the Liouvillian, $\nu \in [0,3]$. (f-i) Corresponding left eigenmatrices. Results obtained using $A=B=\omega = 1, C=0.2$.}
    \label{fig:1_laser}
\end{figure*}

The dynamics of the cavity photons in a laser~\cite{scully1997} can be described by the Liouvillian ($\hbar=1$) \begin{align}
    \mathcal{L}(\rho) = -i[\omega n,\rho \,] & + A \mathcal{D}[a^\dagger](\mathcal{S}^{-1}(\rho)) \nonumber \\
    & + C \mathcal{D}[a](\rho),\label{eq:laser_liouvillian}
\end{align} where $\mathcal{D}[L](\rho) = L\rho L^\dagger - \mathcal{A}[L](\rho)$ is a Lindblad form with jump operator $L$, and $\mathcal{A}[L](\rho) = \{L^\dagger L,\rho\}/2$. This form, with an additional nonlinear interaction term, also describes polariton condensates~\cite{whittaker2009}, atom lasers~\cite{thomsen2002}, and linewidth enhancement~\cite{henry1982}.  The last term in Eq.~\eqref{eq:laser_liouvillian}, proportional to $C$, describes the cavity decay, while the second term, proportional to $A$, describes the gain. Since above threshold gain exceeds loss, $A>C$, one must include gain saturation, captured by the superoperator $\mathcal{S}(\rho) = \rho+\{n,\rho\}B/2A$. It is this nonlinearity which forms the limit cycle, with the mean photon number fixed such that the saturated gain and loss balance
, while the phase remains free. Away from the classical limit quantum fluctuations introduce phase diffusion, leading to a steady-state which is uniformly distributed in phase~\cite{scully1997}. 

In the following we will discuss our results in comparison to semiclassical or mean-field approximations, in which the field operators are replaced with their expectation values, $a\rightarrow \langle a\rangle$. Such an approximation can be obtained for Eq.~\eqref{eq:laser_liouvillian} by replacing the photon number operator in the gain saturation with its average, $S(\rho)=\rho+\{n,\rho\}B/2A\approx\rho+\{|\langle a \rangle|^2,\rho\}B/2A$. This gives the widely used form \begin{equation} \frac{d \langle a\rangle}{dt} =\frac{1}{2}\left(\frac{A}{1+B|\langle a\rangle|^2/A}-C\right)\langle a\rangle. \label{eq:classtheory}\end{equation} The laser threshold is at $A=C$. The steady-state photon number $\langle n\rangle\approx |\langle a\rangle|^2$ is zero below threshold, and \begin{equation} \langle n \rangle \approx \ |\langle a \rangle|^2=A(A-C)/BC\label{eq:mfss}\end{equation} above it.

Rather than studying the steady-state of the Liouvillian, we focus here on its transient dynamics, which retain certain characteristics of the classical limit cycle. We consider its spectrum, obtained from the eigenvalue problem \begin{align}
    &\mathcal{L}\kett{R_{\lambda }} = \lambda \kett{R_{\lambda }} \nonumber \\
    &\mathcal{L}^{\dagger} \kett{L_{\lambda }} = \lambda^* \kett{L_{\lambda }}, \nonumber
\end{align}
where $\kett{R_{\lambda }},\kett{L_{\lambda }}$ are the right and left eigenvectors of the Liouvillian superoperator in Liouville space. Here the double kets denote vectors in Liouville space (equivalent to operators in Hilbert space, $\rho \rightarrow \kett{\rho}$), and the calligraphic symbols denote operators in Liouville space (superoperators in Hilbert space, $\mathcal{L}(\cdot) \rightarrow \mathcal{L}$). The eigenvectors are biorthogonal
\begin{align}
    \bbrakett{L_{\lambda }}{R_{\gamma }} = \delta_{\lambda \gamma}.
\end{align}
This Liouvillian eigenbasis enables the dynamics of the density matrix to be decomposed as 
\begin{align}
    \kett{\rho(t)} =& e^{\mathcal{L}t}\kett{\rho_0} \nonumber \\
    =& \sum_{\lambda } e^{\lambda t} \bbrakett{L_{\lambda }}{\rho_0} \kett{R_{\lambda }},\label{eq:rhodynamics}
\end{align}
where the overlap of the initial state with the left eigenvectors corresponds to the contribution of the corresponding right eigenvectors. The real parts of the non-zero eigenvalues $\lambda$ are all negative and correspond to decay rates, while the imaginary parts correspond to frequencies.

Aspects of the Liouvillian spectra have recently been discussed for the Scully-Lamb laser model~\cite{minganti2021}, the squeezed quantum van der Pol oscillator~\cite{cabot2024}, and a pair of coupled qubits~\cite{cabot2021}, among other models. The eigenspectrum of the Fokker-Planck equation, which approximates the laser Liouvillian, was studied in earlier works~\cite{risken1996}. As pointed out in~\cite{dutta2025}, a key feature of the quantum limit cycle is a band of eigenmodes with small decay rates, which vanish in the thermodynamic limit~\cite{minganti2021}, where the Schawlow-Townes linewidth, due to phase diffusion, is zero. This can be seen in the spectrum of Eq.~\eqref{eq:laser_liouvillian} shown in Fig.~\ref{fig:1_laser}(a). For these parameters the photon number is small, $\langle n\rangle\approx 4$, so there are significant quantum effects. Nonetheless, the classical limit cycle produces a characteristic band of slowly-decaying eigenmodes. It is associated with the phase fluctuations of the field, and separated by a gap from the next band, which includes the intensity fluctuations. 

To simplify the numerical calculations and to better understand the properties of the laser one can use the weak $U(1)$ symmetry~\cite{minganti2021} given by the superoperator $\mathcal{U} = e^{-i\theta \hat{n}} \otimes e^{i\theta \hat{n}}$, which commutes with the Liouvillian, $[\mathcal{L},\mathcal{U}] = 0$. This implies that the eigenmatrices of the Liouvillian must also be eigenmatrices of the symmetry,
\begin{align}
    \mathcal{U}\kett{\psi_{\nu\mu}} = e^{i\nu \theta} \kett{\psi_{\nu j}} ,
\end{align}
where $\nu$ is an integer. It thus constrains the form of the eigenmatrices to
\begin{align}
    \kett{\psi_{\nu j}} = \sum_{m} c_{m,\nu j} \ket{m}\otimes \ket{m-\nu},\label{eq:emodestruc}
\end{align}
which can be used to simplify numerical calculations by restricting them to a particular symmetry block $\nu$. Due to the symmetry, the imaginary part of all eigenvalues of a particular block is given by $\Im[\lambda_{\nu j}] = i\nu\omega$. Ordering the eigenvalues by real part, $\lambda_{\nu 0}$ corresponds to all eigenvalues of the limit cycle parabolic branch, which we focus on in the following. 

Figs.~1(b-e) show the real part of the Husimi Q-functions $Q_{\nu 0}(\alpha) = \bra{\alpha} R_{\nu 0}\ket{\alpha} /\pi$ for the first few right eigenmatrices of the Liouvillian. Note that except for the steady state, for which $\Tr[R_{00}] =1$, the eigenmatrices are traceless and non-Hermitian, and therefore the result is not a quasiprobability distribution. As expected for a laser or limit-cycle the phase-space distribution for the steady-state, $R_{00}$, is peaked at a non-zero intensity, and independent of phase. The remaining phase-space distributions have the form $Q_{\nu j}(\alpha) = Q_{\nu j}(\abs{\alpha})e^{i\nu\arg(\alpha)}$ due to the $U(1)$ symmetry. As can be seen from Figs. 1(b-e) and Eq.~\eqref{eq:rhodynamics}, the parabolic $j=0$ branch can be used to represent a non-uniform phase distribution on the limit cycle as a superposition of modes of different $\nu$. The components with $\nu\neq 0$ then decay as the phase information in the initial state is lost. 

These connections can be further strengthened by considering the conserved quantities of the dynamics. In a classical dynamical system a limit cycle can be understood as an intermediate situation between a fully conservative system, with many conserved quantities, and one with a single fixed point, with none. Specifically, the simplest form of limit cycle has one conserved quantity, corresponding to the phase in the rotating frame~\cite{pikovsky2001, balanov2009}. For an ensemble, in which we associate probabilities with states, this corresponds to conservation of the probability distribution of the phase, $P(\theta)=\sum_{m} e^{im\theta} P_m $, or the equivalent Fourier components $P_m$.

In the classical limit the decay rates of the $j=0$ modes, given by the real parts of the eigenvalues, vanish~\cite{dutta2025,minganti2021} and the parabolic branch lies along the imaginary axis, breaking the $U(1)$ symmetry of the steady state \cite{minganti2018a}. This then gives an infinite number of conserved quantities, which are given by the corresponding left eigenvectors of the Liouvillian~\cite{albert2014,albert2018,buca2022,blume-kohout2010}, since then
\begin{align}
    \Tr[  e^{-i\nu\omega t} L_{\nu 0} \rho(t)] =& \bbrakett{L_{\nu 0 }}{\rho_0}  
\end{align} is constant. Away from the classical limit, the decay rates are non-zero, and these become quasi-conserved quantities. Their phase-space representation is shown in Fig.~\ref{fig:1_laser}(f-i), and can be seen to approximate the phase operator, with $Q$ function
\begin{align}
    Q_{\nu j}(\alpha) \approx&  e^{i \nu \arg(\alpha)},
\end{align}
up to normalization. 

These numerical results are consistent with an analytical calculation of the $Q$-functions of the eigenmodes given in Appendix~\ref{appendix:single_laser}. In the semiclassical regime $\abs{\alpha}\gg 1$, for $B=A$, we find
\begin{align}
    Q_{\nu j}(\alpha) \approx & e^{-\nu^2 /4 \abs{\alpha}^2} \abs{\alpha}^{C\nu^2/2A}e^{i \nu \arg(\alpha)},
\end{align}
which approaches the phase operator well above threshold, $A \gg C$. The eigenvalues are \begin{equation}
\lambda_{\nu,j} = -jC -\frac{\nu^2 C^2}{4A},\label{eq:anespec}
\end{equation} which is similar to the result for a spin model in~\cite{dutta2025} and agrees with results for the laser from other semiclassical methods~\cite{risken1996,scully1997}.

\begin{figure}
\includegraphics[width=\columnwidth]{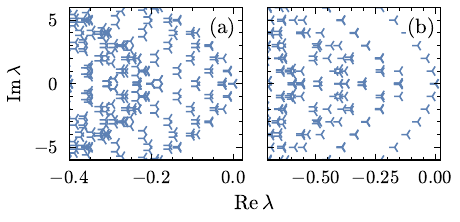}\caption{Spectrum of the Liouvillian for two lasers with detuning $\delta\omega=0.1$, with no coupling (a), and with dissipative coupling $g=0.2$. Results obtained using $A=B=\omega = 1, C=0.2$. \label{fig:dissipative_spectrum_2}}
\end{figure}

The $U(1)$ symmetry can be used to connect the eigenspectrum to the experimentally observable coherence functions of the emission. Eq.~\eqref{eq:emodestruc} implies that the steady-state correlation functions of operators which do not change the photon number, notably the intensity $I(t)=\langle a^\dagger(t) a(t)\rangle$ and the second-order coherence $g^{(2)}(\tau)=\langle I(t+\tau)I(t)\rangle/\langle I(t)\rangle^2$, are determined by the $\nu=0$ sector of the Liouvillian. Eq.~\eqref{eq:anespec} then gives the long-time decay of intensity fluctuations and correlations at the expected rate $-\lambda_{0,1}=C$, in the semiclassical region well above threshold (see~\cite{whittaker2009} for a treatment of the threshold region). Similarly, correlation functions of operators which change the photon number by one, notably the first-order coherence $g^{(1)}(\tau)=\langle a^\dagger (t)a(t+\tau)\rangle$, probe the $\nu=\pm 1$ sectors, and have long-time decay rate $-\lambda_{1,0}=C^2/(4A)$. Since the Fourier transform of $|g^{(1)}(\tau)|$ is the emission spectrum this is the linewidth, up to numerical factors, and agrees with the standard Schawlow-Townes result. More generally, Eqs.~\eqref{eq:emodestruc} and \eqref{eq:anespec} imply that the decay rate of the $n^{th}$ order correlation function $\langle [a^\dagger(t)]^n [a(t+\tau)]^n\rangle$ is $n^2 C^2/(4A)$. Thus the $n$-photon spectra are $n^2$ times broader than the usual 1-photon ones. This could provide an experimental signature of the parabolicity of the limit-cycle branch, which is a quantum effect absent in the classical limit~\cite{dutta2025}.

It is useful to note that the classical and quantum systems differ in two important respects. One, emphasized above, is that the conservation of the Fourier components of the phase distribution in the classical limit is replaced by a quasi-conservation. The second is that classical systems have an infinite number of such Fourier components. The conserved probability distribution of the phase is then that of a continuous variable, and encodes an infinite number of bits. Quantum systems, however, often have a finite number of states, in which case the probability distribution of the phase is specified by a finite number of Fourier components, and encodes only a finite number of bits. For both reasons reproducing classical limit-cycle behavior in quantum systems requires a large Hilbert space, such as that of bosons \cite{lee2013,walter2014,walter2015,chia2020,benarosh2021} or a large spin \cite{dutta2025}. Time-dependent steady states can be achieved in quantum systems with small Hilbert spaces, but they typically rely on engineering decoherence free subspaces, where a set of dark states do not undergo dissipation \cite{lidar1998,kempe2001}. Unlike the physics of limit cycles, this generally requires specific models, often involving strong symmetries~\cite{albert2014} or fine-tuning of parameters.

\section{Coupled lasers}\label{sec:coupled_lasers}

In the following we consider the coupling of two detuned but otherwise identical lasers, described by two copies of the Liouvillian in Eq.~\eqref{eq:laser_liouvillian} with frequencies $\omega_{12} = \omega_0 \pm \delta \omega/2$. We consider both reactive and dissipative couplings~\cite{ding2019}. Reactive couplings are those arising from a Hamiltonian, in particular the tunneling coupling, $H_{c}=J(e^{i\phi_J} a_1^\dagger a_2 + e^{-i\phi_J}a_2^\dagger a_1)$ . In lattices of polariton condensates~\cite{moroney2021}, for example, this describes the Josephson coupling between neighboring lattice sites, due to the overlap of wavefunctions; for lasers, such terms arise if one considers laser modes which form within a single cavity structure. 

A different form of coupling, commonly used to achieve synchronization in lasers~\cite{vladimirov2003}, involves taking the output of one and feeding it optically into another, and vice versa. This is a dissipative coupling~\cite{aleiner2012a}, mediated by the environment, whose effects are modified by the feedback. Dissipative couplings also occur in lattices of polariton condensates, in geometries where polaritons can flow from one lattice site to another~\cite{kalinin2019}. We consider the simplest case where an equal fraction of the output of each emitter (laser or condensate) is fed into the other, with zero phase shift for the propagation. For this case the contributions to the equations-of-motion from the loss terms, dissipative coupling, and reactive coupling, are \begin{align} \dot a_1 &= -(C+g)a_1/2 + g a_2/2 -iJ e^{i\phi_J} a_2, \nonumber \\ \dot a_2 &= -(C+g) a_2/2 + g a_1/2-iJe^{-i\phi_J}a_1.\label{eq:coupledeoms}\end{align} Diagonalizing this system we see that the dissipative coupling is equivalent to an additional non-local Lindblad dissipator which, along with the reactive coupling, leads to the interaction Liouvillian
\begin{align}
    \mathcal{L}_I(\rho) = g\mathcal{D}[a_1-a_2](\rho) -i[H_c,\rho ].\label{eq:interaction_liouvillian}
\end{align}

The interaction Liouvillian does not commute with the local $U(1)$ symmetry discussed above, but has a collective $U(1)$ symmetry, given by $\mathcal{U} = e^{-i\theta (\hat{n}_1+\hat{n}_2)} \otimes e^{i\theta (\hat{n}_1+\hat{n}_2)}$, such that the eigenmatrices of the symmetry and Liouvillian operators are now of the form
\begin{align}
    \kett{\psi_{\nu j}} = \sum_{mnp} c_{mpn,\nu j} \ket{m,p}\otimes \ket{n,m+p-n-\nu},
\end{align}
which can again be exploited in order to block-diagonalize the Liouvillian. The imaginary part of the eigenvalues is no longer fixed by symmetry, but for small detuning and coupling, the least damped modes have $\Im[\lambda_{\nu j}] \approx i\nu \omega_0$.

Similarly to the single-laser case, the symmetry implies that the dynamics of particular observables is controlled by particular symmetry blocks. Each block now corresponds to a different total number of excitations $n_1 + n_2$, and the symmetry sector $\nu$ determines the dynamics of operators, such as $a_1^{\dagger \, p}a_2^{\dagger \,q}a_1^{r}a_2^s$, which change the excitation number by $\nu=p+q-r-s$. Thus the first order coherence functions, $\langle a_i^\dagger(t)a_j(t+\tau)\rangle$, are again determined by $\nu=\pm 1$, while $\nu=0$ describes the intensity correlations, $\langle a_i^\dagger (t+\tau) a_j (t+\tau) a_k^\dagger (t) a_l (t)\rangle$. The $\nu=2$ block gives the dynamics of two-photon correlations, encoded in coherence functions of the form $\langle a_i^\dagger(0)a_j^\dagger(0)a_k(\tau)a_l(\tau)\rangle$.

\subsection{Dissipative coupling and synchronization}\label{sec:dissipative}

In the semiclassical limit, the dissipative coupling of lasers produces a synchronization transition described by a two-oscillator Kuramoto model~\cite{vladimirov2003} or Adler equation. Above threshold the laser amplitude is fixed by the gain and loss, and we may write $\langle a_i\rangle=\sqrt{\rho_i}e^{i\theta_i}$. The equations-of-motion for the phases, 
\begin{align}
\dot\theta_1=-\omega_1+\sqrt{\frac{\rho_2}{\rho_1}}\left[\frac{g}{2}\sin(\theta_2-\theta_1)-J\cos(\theta_2-\theta_1-\phi_J)\right], \nonumber\\ \dot\theta_2=-\omega_2+\sqrt{\frac{\rho_1}{\rho_2}}\left[\frac{g}{2}\sin(\theta_1-\theta_2)-J\cos(\theta_2-\theta_1-\phi_J)\right], 
\end{align} 
contain sine (cosine) terms from the dissipative (reactive) couplings in Eq.~\eqref{eq:coupledeoms}. When $\rho_1=\rho_2$ the phase difference $\phi=\theta_2-\theta_1$ obeys the Adler equation \begin{equation}\label{eq:adler} \dot\phi=\delta\omega-g\sin(\phi),\end{equation} which contains no contribution from the reactive coupling. It therefore cannot produce synchronization with $\rho_1=\rho_2$, unlike the dissipative coupling, which does so within the usual Arnold tongue \begin{equation}|\delta\omega|<g.\label{eq:arnold}\end{equation}

\begin{figure}
\includegraphics[width=0.8\columnwidth]{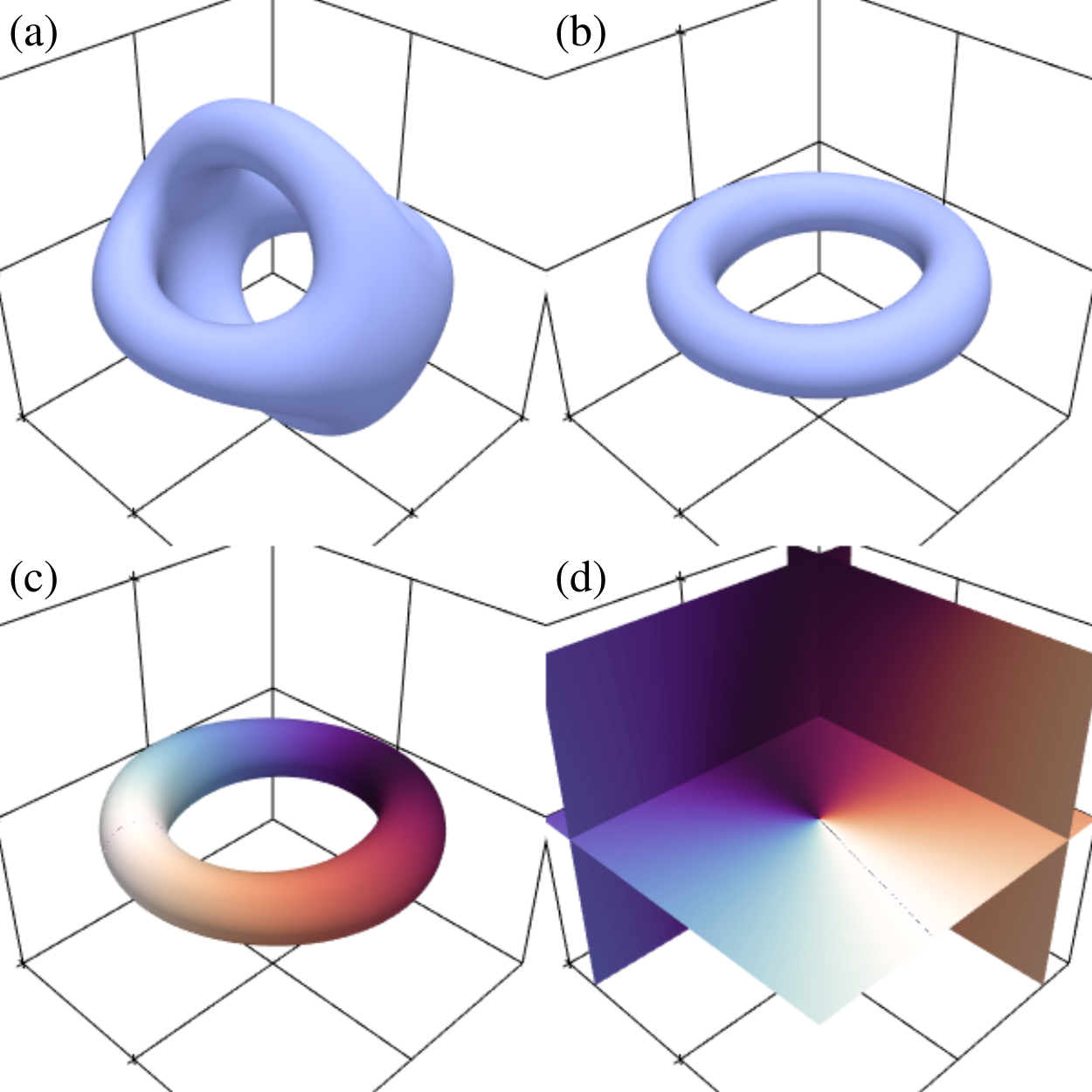}
\caption{Phase-space distributions of the Liouvillian eigenmatrices for two lasers, given by the Q-functions in the basis of collective modes (see text), $Q(\beta_1=x+iy,\beta_2=z)$. The z-axis is vertical and all axes span $[-4,4]$. (a,b) Isosurfaces of the right eigenmatrix for $\lambda=0$, $R_{00}$, for uncoupled desynchronized lasers (a), and coupled synchronized lasers (b). All isosurfaces plotted at $2\max(Q)/3$. (c) Magnitude isosurface for the right eigenmatrix of the slowest $\nu=1$ mode, $R_{10}$, for synchronized lasers, colored by $\arg Q$. (d) $\arg Q$ for the corresponding left eigenmatrix, $L_{10}$, on planes in phase-space. $\phi=0$ in (a) and $0.50$ in (b,c,d).\label{fig:husimiqcoupled} }
\end{figure}

Similarly to the limit cycle, we can look at the Liouvillian spectrum to understand how synchronization emerges from quantum dynamics. Fig.~\ref{fig:dissipative_spectrum_2} compares the spectra for two detuned oscillators with and without coupling. In the case without coupling, Fig.~\ref{fig:dissipative_spectrum_2}(a), the slowest modes form a parabolic branch, which now consists of doublets that are split in frequency (the imaginary part) by the detuning for odd $\nu$. This branch is also no longer clearly separated from the others. Fig.~\ref{fig:dissipative_spectrum_2}(b) shows a spectrum above the synchronization threshold, Eq.~\eqref{eq:arnold}. Here we see a well-isolated parabolic branch, very similar to that in the single-oscillator case.

To show the form of the eigenmatrices it is useful to transform to collective modes 
\begin{align}
    &b_1 = \frac{1}{\sqrt{2}}(a_1+e^{i\phi}a_2), \nonumber \\
    &b_2 = \frac{1}{\sqrt{2}}(e^{-i\phi}a_1-a_2).
\end{align}
Fig.~\ref{fig:husimiqcoupled} shows the Husimi functions, $Q(\beta_1,\beta_2)=\langle \beta_1,\beta_2|(L,R)_{\nu0}|\beta_1,\beta_2\rangle/\pi^2$, for the eigenmatrices corresponding to some eigenvalues in Fig.~\ref{fig:dissipative_spectrum_2}. The collective $U(1)$ symmetry results in $Q(\beta_1,\beta_2) = Q(\beta_1,\abs{\beta_2})e^{i\nu\arg(\beta_2)}$ (see Appendix~\ref{appendix:sym}). Therefore it is enough to take $\beta_2$ to be real and show $Q$ in the three-dimensional subspace $\beta_1=x+iy,\beta_2=z$. For the steady-state of the uncoupled case we set $\phi=0$, giving the result in Fig.~\ref{fig:husimiqcoupled}(a) which shows the two limit cycles in the two collective modes. For the synchronized case we expect the steady-state to show a single limit cycle, in a collective mode with phase difference close to the classical result, given by Eq.~\eqref{eq:adler}, $\phi=\sin^{-1}(\delta\omega/g)$. We find numerically that the maximum of the $Q$ function is at $\phi=0.50\approx \arcsin(1/2)=\pi/6$. Plotting the steady-state $Q$ function using this $\phi$, in Fig.~\ref{fig:husimiqcoupled}(b), indeed shows a single limit cycle, with the state distributed over a ring in phase-space, $|\beta_1|\neq 0$, that lies in the plane $\beta_2=0$. 

Figs.~\ref{fig:husimiqcoupled}(c) and (d) show the right- and left- eigenmatrices for slowest $\nu=1$ mode, $R_{10}$, $L_{10}$, in the synchronized regime. Similarly to the single-laser case the distribution of the right eigenmatrix, $R_{10}$, follows that of the steady-state, $R_{00}$, in magnitude, but incorporates an additional phase winding. The left eigenmatrix $L_{10}$ gives the corresponding conserved quantity. It is uniform in magnitude (not shown), while its argument winds with $\beta_1$ as shown in Fig.~\ref{fig:husimiqcoupled}(d). It is thus the phase operator for the limit cycle, as in the single-laser case, and the slow decay of this mode corresponds to quasi-conservation of the lowest Fourier component of the distribution of phase around the limit cycle. Note that for $g \gg C$ the eigenvalues characterizing the decay of the $\beta_2$ mode get pushed to more negative values, and the spectrum in Fig.~\ref{fig:dissipative_spectrum_2}(b) approaches the simpler form in Fig.~\ref{fig:1_laser}(a).

\begin{figure}
    \centering
\includegraphics[width=\columnwidth]{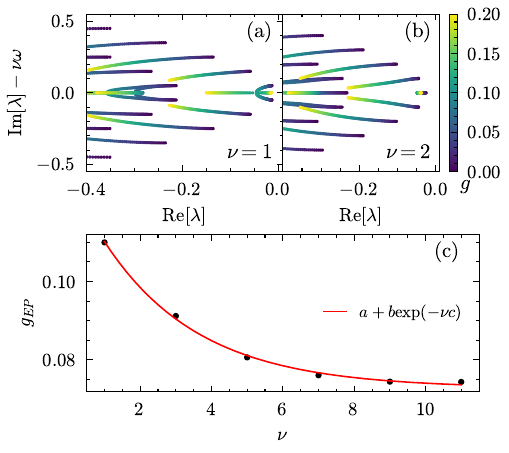}
    \caption{(a-b) Evolution of the spectrum of Liouvillian blocks $\nu=1,2$ with different dissipative coupling strength $g$. $\nu=1$ displays an exceptional point, absent in $\nu=2$. Using parameters $A=B=1, C = 0.2, \delta\omega = 0.1$. (c) Coupling at which the EP of the two leading eigenvalues occurs for different odd harmonics $\nu$, displaying an exponential dependence.}
    \label{fig:dissipative_evolution}
\end{figure}

To understand the transition to synchronization we focus on the eigenvalues of the symmetry blocks $\nu=1$, characterizing the dynamics of $\expval{a_j}$ and the first-order coherence, and $\nu =2$. We show how these eigenvalues evolve as the dissipative coupling increases in Fig.~\ref{fig:dissipative_evolution}(a,b). For $\nu =1$ the two slowest eigenstates correspond, initially, to the tensor product of two uncoupled eigenmatrices, $R_{10} \otimes R_{00}$ and  $R_{00} \otimes R_{10}$, describing a phase mode excitation in the first and second oscillator respectively. The eigenvalues have degenerate real parts, and an imaginary gap, i.e., frequency splitting, $\delta\omega$. As the coupling is turned on the imaginary gap reduces until the eigenvalues meet at an exceptional point (EP), signaling the transition to the synchronized regime. From this point on, both eigenvalues remain at the same imaginary part, i.e., they have the same frequency, and the real gap between them increases such that only one of the eigenvalues contributes significantly to the long-time dynamics. 

This behavior can be compared with that of semiclassical treatments of lasers~\cite{ding2019} where EPs have been considered in a variety of models and regimes~\cite{ji2023}, and is connected to the generalized EPs recently discussed in the dynamics of certain non-linear classical systems~\cite{fruchart2021,weis2025}. The semiclassical approximation here is Eqs.~\eqref{eq:classtheory} and \eqref{eq:coupledeoms}. Linearizing these equations about the mean-field steady-state, Eq.~\eqref{eq:mfss}, one finds that the fluctuations obey \begin{align}
    \frac{d}{dt}\mqty(\delta\expval{a_1} \\ \delta\expval{a_2}) = \frac{1}{2}\mqty(i(2\omega_0+\delta\omega)  & g \\ g & i(2\omega_0-\delta \omega))\mqty(\delta\expval{ a_1} \\ \delta\expval{a_2}).\label{eq:lineom}
\end{align} This linear system has an exceptional point at $g=g_{c1}= \delta\omega$, in agreement with the classical synchronization threshold, Eq.\ \eqref{eq:arnold}. 

The $\nu=1$ EP in Fig.~\ref{fig:dissipative_evolution}(a) lies close to the classical result, but is shifted slightly by quantum effects. More profound differences between the classical and quantum theories appear when one considers the other symmetry sectors. In the classical theory all expectation values are assumed to factorize. Thus the dynamics are entirely characterized by the equations-of-motion ~\eqref{eq:lineom} and the EP at $g_{c1}$. This is not the case for the quantum theory, as we see in Fig.~\ref{fig:dissipative_evolution}(b), which shows the evolution of the eigenvalues in the $\nu=2$ sector, and (c), which shows the positions of the EPs in different symmetry sectors. 

As shown by the example of $\nu=2$ in Fig.~\ref{fig:dissipative_evolution}(b), there are no EPs in the slow modes for even $\nu$. Instead there is a single eigenvalue with highest real part that remains isolated throughout the synchronization transition. This can be understood in terms of the three slowest modes present for $g=0$. The slowest of these three is a single mode, with eigenmatrix $R_{10}\otimes R_{10}$, in which there is one phase excitation in each oscillator. The other two form a pair with equal decay rates. They correspond to two phase excitations in the first oscillator and none in the second, $R_{20}\otimes R_{00}$, and vice versa. These eigenvalues have different imaginary parts due to the detuning, however, they move out of the slow part of the spectrum as the coupling increases.

For higher odd symmetry sectors the least damped eigenmodes are always a pair of eigenvalues that undergo an EP, similarly to the behavior for $\nu=1$. However, contrary to the classical solution where all harmonics undergo an EP at the same point, we observe that different odd moments of the phase distribution synchronize at different values of the coupling, resulting in a cascade of EPs in the different symmetry sectors. Higher harmonics synchronize at lower values of the coupling. As shown in Fig.~\ref{fig:dissipative_evolution}(c), this can be fitted by an exponential decay, implying all EPs occur within a finite coupling range which tends to zero in the classical limit.

\begin{figure}[t]
    \centering
\includegraphics[width=\linewidth]{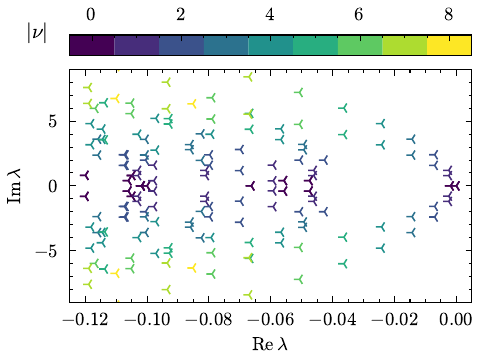}
    \caption{(a) Liouvillian spectrum for reactively coupled oscillators in the classical bistable regime, showing two branches of eigenvalues gapped from the bulk. Computed with $A=B=\omega_0 = 1, C=0.1, \delta\omega = 0.1, J=2\delta\omega=0.2, g=0$.}
    \label{fig:reactive_spectrum}
\end{figure}

\begin{figure}
\centering
\includegraphics[width=0.8\linewidth]{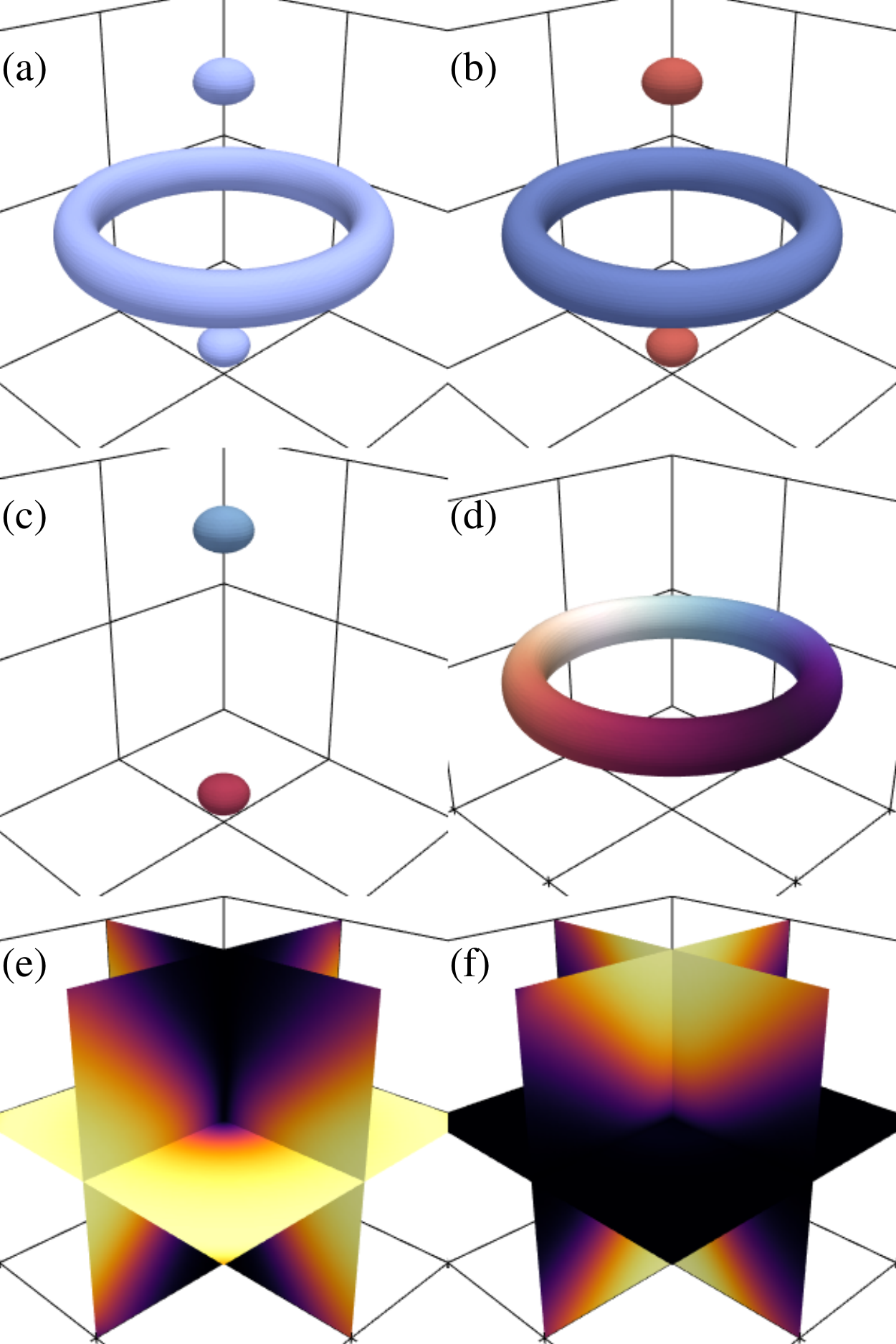}
\caption{Phase-space distributions for the eigenmatrices of the Liouvillain with reactive coupling $J=2\delta\omega$, given by the Q-functions in the basis of collective modes, $Q(\beta_1=x+iy,\beta_2=z)$ (see text). The z-axis is vertical and all axes span $[-5.2,5.2]$. (a) Steady-state, $R_{00}$. (b) Slow mode in the $\nu=0$ sector, $R_{01}$. (c,d) Slow modes in the $\nu=1$ sector, $R_{10}$ and $R_{11}$. All plots show isosurfaces of $|Q|$, colored in (b,c,d) according to $\arg Q$. (e,f) Magnitude of $Q$ for the two slowest left modes in the $\nu=1$ sector, $L_{10}$ and $L_{11}$. $A=B=\omega_0=1, C=0.1, \delta\omega=0.1, J=0.2, g=0$.  \label{fig:reactiveqfuns}}
\end{figure}

\subsubsection{Role of PT symmetry}
The exceptional points are protected by a Liouvillian PT symmetry \cite{prosen2012,nakanishi2026,nakanishi2026a} in each of the symmetry blocks. Consider the parity operator that swaps both lasers, $P a_{1(2)}P = a_{2(1)}$, and the time-reversal operator given by the conjugation operator, $Ti=-iT$ . Since both lasers are identical, apart from the detuning, the single laser dissipators are symmetric under parity, as is the chosen dissipative coupling. All dissipators are also symmetric under time-reversal due to the jump operators being real. Regarding the Hamiltonian, we can divide it in a global term $H_0 = \omega_0 (n_1 + n_2)$ and the detuning $H_d = (\delta\omega/2)(n_1-n_2)$. The Liouvillian term corresponding to $H_0$ commutes under parity, $P \mathcal{L}_{H_0} P = \mathcal{L}_{H_0}$, and anticommutes under time-reversal, $T \mathcal{L}_{H_0} T = - \mathcal{L}_{H_0}$, while the detuning term anti-commutes under both parity, $P \mathcal{L}_{H_d} P = - \mathcal{L}_{H_d}$ and time-reversal, $T \mathcal{L}_{H_d} T = -\mathcal{L}_{H_d}$. The combined action leads to  
\begin{align}
    TP(\mathcal{L}_{H_0}+\mathcal{L}') PT = -\mathcal{L}_{H_0}+\mathcal{L}',
\end{align}
where $\mathcal{L}'$ accounts for all other terms in the Liouvillian. When restricted to each symmetry block, the global term commutes with the rest of the Liouvillian, since in this subspace it is proportional to the identity, $\mathcal{L}_{H_0} = i\nu\omega_0 \mathbb{I}$. Consider an eigenstate of the Liouvillian from symmetry block $\nu$, with eigenvalue $(\mathcal{L}_{H_0}+\mathcal{L}')\ket{\lambda_{\nu j}} = (i\nu\omega + \delta\lambda_{\nu j}) \ket{\lambda_{\nu j}}$. Under the PT symmetry we have
\begin{align}
    \mathcal{L} PT \ket{\lambda_{\nu j}} =& PTTP \mathcal{L} PT \ket{\lambda_{\nu j}} \nonumber \\
    =& PT(-\mathcal{L}_{H_0}+\mathcal{L}')\ket{\lambda_{\nu j}} \nonumber \\
    =&PT(-i\nu\omega +\delta\lambda_{\nu j} )\ket{\lambda_{\nu j}} \nonumber \\
    =& (i\nu\omega +\delta \lambda_{\nu j}^*)PT \ket{\lambda_{\nu j}},
\end{align}
which leads to the symmetry around the $\Im[\lambda] = i\nu\omega_0$ line observed in the spectrum in Fig.~\ref{fig:dissipative_evolution}(a,b). For each of the $\nu$ symmetry blocks we can therefore define a \emph{broken} PT symmetric phase where the two slowest modes are related by $PT\ket{\psi_1} = \ket{\psi_2}$, corresponding to the desynchronized regime, and an \emph{unbroken} PT symmetric phase where the slowest mode is PT symmetric, $PT\ket{\psi} = \ket{\psi}$, corresponding to the synchronized regime.

\subsection{Reactive coupling and bistability}\label{sec:reactive}

\begin{figure}[t]
    \centering
\includegraphics[width=\linewidth]{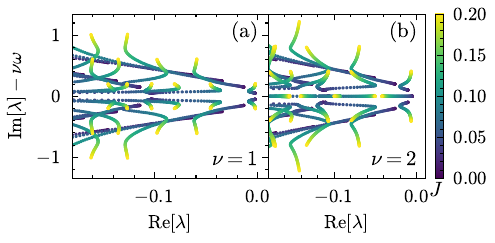}
    \caption{Evolution of the Liouvillian spectrum for the (a) $\nu =1$ and (b) $\nu=2$ symmetry blocks as reactive coupling is increased. Computed with $A=B=\omega_0 = 1, C=0.1, \delta\omega = 0.1, J=2\delta\omega=0.2, g=0$.}
    \label{fig:reactive_evolution}
\end{figure}

As noted above, purely reactive coupling cannot drive synchronization between two lasers when $\rho_1=\rho_2$. However, in the classical limit transitions do occur between lasing at multiple frequencies and at a single frequency~\cite{ding2019}, which can be understood in terms of mode competition~\cite{eastham2008}. The relevance of mode competition can be seen by eliminating the reactive coupling using the transformation \begin{equation}\begin{pmatrix}b_1 \\ b_2 \end{pmatrix}=\begin{pmatrix} \cos(\theta) & \sin(\theta)e^{i\phi} \\ \sin(\theta)e^{-i\phi} & -\cos(\theta)\end{pmatrix}\begin{pmatrix} a_1 \\ a_2 \end{pmatrix}\label{eq:mixtran}\end{equation} with \begin{equation}\tan 2\theta=-2J/\delta\omega,\label{eq:mixangle}\end{equation} and $\phi=\phi_J$. This gives collective modes which for large detuning $|\delta\omega|\gg J$ are almost unchanged from the original ones. Since the nonlinear gain is local, this leads to independent lasing in the two modes, at their two different frequencies. For small detuning, however, the modes become superpositions. This gives rise to cross-mode terms in the nonlinear gain, in which the occupation of one collective mode reduces the gain for the other. Such coupling via the nonlinear gain can permit lasing in only one of the two collective modes and bistability~\cite{eastham2008}, where there are two stable limit cycles with different basins of attraction. The two cycles correspond to phase-locking and anti-locking in the original basis, and present an asymmetry of the amplitudes and different frequencies~\cite{eastham2008,walter2015,ding2019}. In the quantum regime, much less studied, the existence of multiple peaks in the phase-phase correlations of the steady state has been interpreted as a consequence of bistability~\cite{lee2013,thomas2026}. 

In Fig.~\ref{fig:reactive_spectrum}(a) we show the Liouvillian spectrum of the reactively coupled lasers in the bistable regime. Note that when $g=0$, as here, the spectrum does not depend on the phase of the reactive coupling as it can be  removed by a gauge transformation, $a_1 \rightarrow a_1 e^{i\phi_J}$. Similarly to the synchronized state (Fig.~\ref{fig:dissipative_spectrum_2}(b)) we see a gap separating a slow set of eigenvalues from the bulk. However, this spectrum now contains two slow branches. For $\nu\neq 0$, the slow modes all form doublets, with equal real parts and different imaginary parts. This differs from the uncoupled case (Fig.~\ref{fig:dissipative_spectrum_2}(a)), where the slowest modes alternate between doublets, for odd $\nu$, and triplets, for even $\nu$. In addition, the slow part of the spectrum for $\nu=0$ contains both the steady state and a second nearby mode on the real axis. This too differs from the uncoupled case, where the slowest $\nu=0$ modes are a doublet lying deeper in the spectrum.

To explore how this spectrum relates to the classical bistable solution we look to the eigenmatrices of the least damped modes. In Fig.~\ref{fig:reactiveqfuns} we show the phase-space distributions for some of the slow modes. These are shown after transforming to the collective modes given by Eq.~\eqref{eq:mixtran}, with the angles $\theta$ and $\phi$ set to correspond to the maxima of the $Q$-function for the steady state. For these parameters we find $\theta=0.67$, and $\phi=0.07$, in agreement with the results of the classical approximation and Eq.~\eqref{eq:mixangle}. With this transformation, the steady-state shown in Fig.~\ref{fig:reactiveqfuns} can be seen to be a distributed between two limit cycles. One of these corresponds to a limit cycle in the first collective mode, occupying the region around $|\beta_1|\neq 0,\beta_2=0$, and the other in the orthogonal mode, occupying the regions around $|\beta_2|\neq 0,\beta_1=0$. The presence of both contributions together is a consequence of the quantum fluctuations, which produce transitions between the two bistable classical solutions, giving a steady state which is an equal mixture of the two. Fig.~\ref{fig:reactiveqfuns}(b) shows the slowest $\nu=0$ mode, for which the Q-function takes the same form as that for the steady state, but with a sign difference separating the regions of the two limit cycles. Thus by combining this mode and the steady state we can represent the density matrix for a system definitely in one or the other of the classical bistable states, which then decays with rate $\lambda_{01}$.

Figures~\ref{fig:reactiveqfuns}(c,d) show the Q-functions for the slowest $\nu=1$ modes. We see that each of these corresponds to one of the two limit cycles that appear in the steady-state, with an additional phase winding. They can thus be used to describe states in which the lowest harmonic of the phase distribution, for one or the other limit cycle, is excited. Similarly to the single-oscillator case, we find that higher slow modes, $\nu>1$, describe the higher harmonics of the phase distribution, again generalized to a situation with two limit cycles. Figs.~\ref{fig:reactiveqfuns}(e,f) show the magnitude of the Q-function for the two slowest left eigenmodes for $\nu=1$. For the most part both eigenmodes occupy different regions of the phase distribution which is the reason bistability emerges, as a total separation in phase space of the two left eigenmodes corresponds to two well-defined basins of attraction in the classical dynamics. Localized initial density matrices will overlap mostly with one of the two left eigenmodes, unless they are near the boundary, and therefore the evolution will follow one of the two limit cycles. 

To see how this bistability emerges we again study the eigenvalues of $\nu=1,2$ as the reactive coupling is turned on, shown in Fig.~\ref{fig:reactive_evolution}(a-b). In the $\nu=1$ case we find that the leading eigenvalue pair is not much affected. While some of the other eigenvalues acquire larger dissipation, leading to an overall gap with the bulk, other eigenvalues become less damped which results in the gap being much smaller than in the dissipative case. In the $\nu=2$ case, there is a single least damped eigenvalue in the uncoupled case, as mentioned above. When the coupling is turned on this eigenvalue acquires larger damping such that the following pair of eigenvalues becomes the leading ones, resulting in the two branches. Similar behavior is found for the rest of odd and even symmetry blocks. 

Contrary to the dissipative case, there are no singularities (EPs) in the slow modes, and the spectrum always comprises two eigenvalue branches with different frequencies. Thus it does not explain the apparent synchronization in the classical dynamics, where the steady-states each contain a single frequency. In the classical dynamics, however, there are two steady-states, corresponding to two limit cycles with different basins of attraction. This is consistent with the existence of two branches of eigenvalues, rather than one, which contain an additional bit of information corresponding to which of the basins the system is in. As shown by the phase-space distributions in Fig.~\ref{fig:reactiveqfuns}, this is indeed the correct interpretation of the two-branched spectrum, with the two limit cycles corresponding to lasing in either of the two orthogonal modes. This structure implies that there is emission (given by the first-order coherence) at two frequencies, not one, as the system transitions between the two bistable states under the noise. Emission at a single frequency would be observable if the system is prepared in one of the two bistable states, and the spectrum measured over a timescale short compared with the transition rate $\lambda_{01}$. It may be noted that the two limit cycles correspond to approximate phase-locking and anti-locking in the original basis, as previously seen in the phase-phase correlations of reactively-coupled quantum oscillators~\cite{lee2013,thomas2026}. 

\subsubsection{Role of PT symmetry}

\begin{figure}
    \centering
\includegraphics[width=0.8\linewidth]{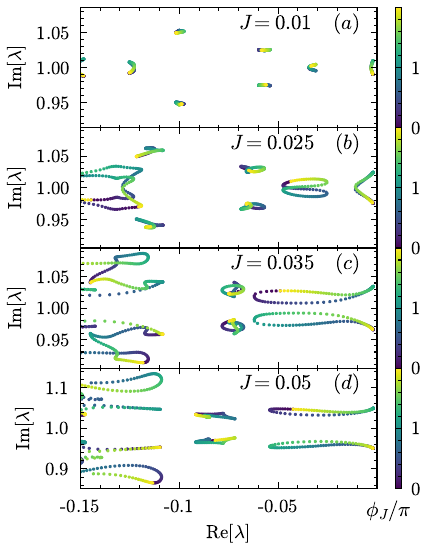}
    \caption{Evolution of the spectrum for $\nu=1$ in the synchronized case $g=2\delta\omega=0.05$, for different values of the reactive coupling $J$, plotted as the phase runs periodically, leading to eigenvalue loops that merge at moderate $J$. Computed with $A=B=\omega_0 = 1, C=0.1$.
    \label{fig:fig7}}
\end{figure}

As in the dissipative case, the spectrum for each $\nu$ is symmetric around the line $\Im[\lambda] = i\nu\omega$, due to the presence of a PT symmetry. While the $PT$ operator anticommutes with the reactive coupling term
\begin{align}\label{eq:sign}
    TP \mathcal{L}_{H_J}PT = -\mathcal{L}_{H_J},
\end{align}
the sign can be absorbed into the gauge transformation $S^\dagger a_2 S = e^{i\pi} a_2$, such that
\begin{align}
    STP \mathcal{L}_{H_J}PTS = \mathcal{L}_{H_J}.
\end{align} Given the absence of exceptional points with varying $J$ (Fig.~\ref{fig:reactive_evolution}), the symmetry therefore implies that the slow $\nu \neq 0$ modes are broken PT-symmetric doublets corresponding to bistability. Furthermore, the eigenmatrices of these pairs are related by $\ket{R_{\nu 0}} = PTS \ket{R_{\nu 1}}$, which transform between the $\beta_1$ and $\beta_2$ coordinates, 
\begin{align}
    S^\dagger TP \begin{pmatrix} b_1 \\ b_2 \end{pmatrix} PTS = \begin{pmatrix} b_2 \\ -b_1 \end{pmatrix},
\end{align} explaining the symmetry that can be inferred between Figs.~\ref{fig:reactiveqfuns}(c) and (d). In the $\nu=0$ sector the first two eigenvalues are real and non-degenerate in the bistable regime, and hence the eigenmatrices are individually eigenmatrices of $PTS$. They are thus either symmetric (eigenvalue +1) or antisymmetric (-1) under the exchange $\beta_1\leftrightarrow\beta_2$, which can be seen to be the case for Figs.~\ref{fig:reactive_spectrum}(a) and (b), respectively.

\subsection{Combined couplings: spectral flow from synchronization to bistability}\label{sec:spectral_flow}

We now consider the case where both couplings are present, where the phase in the reactive coupling can no longer be gauged away. We have seen how PT symmetry plays a fundamental role in the behavior of the system for both purely dissipative and reactive couplings. However, this symmetry is broken when both couplings are present, because the extra sign in \eqref{eq:sign} can no longer be gauged away. Instead, the action of the PT operator on the full Liouvillian leads to a generalized symmetry
\begin{align}
    TP(\mathcal{L}_{H_0}+\mathcal{L}'(\phi_J))PT = -\mathcal{L}_{H_0} + \mathcal{L}'(\phi_J + \pi),
\end{align} which relates the spectrum with coupling phase $\phi_J$ to that with phase $\phi_J+\pi$.  While this symmetry can no longer pin the eigenvalues, it still leads to interesting topological phenomena when we consider how the eigenvalues evolve with the phase. In Fig.~\ref{fig:fig7} we show such case for a fixed dissipative coupling in the synchronized regime, $g>\delta\omega$, and increasing values of the reactive coupling. With a small reactive perturbation, in Fig.~\ref{fig:fig7}(a), the eigenvalues, $\lambda(\phi_J)$, form loops as the phase runs periodically but we are still in a synchronized regime with a single dominant eigenvalue. As mentioned above, we observe that the PT symmetry is broken at any particular value of the phase. Nonetheless, the generalized PT symmetry results in the loops spanned by each eigenvalue being symmetric around the $\Im[\lambda] = i\nu\omega$ line. With weak reactive coupling, there is an unbroken symmetry regime where the two dominant eigenvalue loops map to themselves by the symmetry, $PT\ket{\lambda(\phi_J)} = \ket{\lambda(\phi_J+\pi)}$ and have a real gap between them, implying synchronization. As the reactive coupling is increased the two dominant eigenvalues go through a transition resulting in a broken symmetry regime where the two eigenvalue loops map to each other, $PT\ket{\lambda(\phi_J)} = \ket{\lambda'(\phi_J+\pi)}$. This transition can only happen when the two loops meet at an exceptional point.

\begin{figure}[t]
    \centering
\includegraphics[width=0.8\columnwidth]{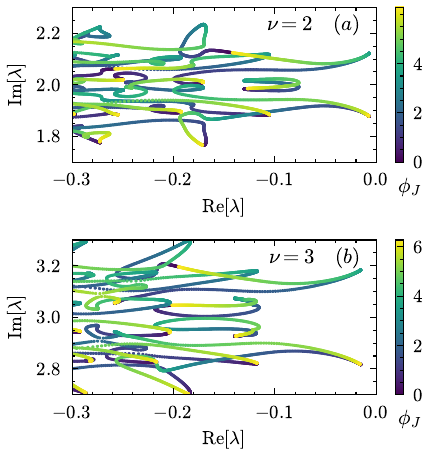}
    \caption{Evolution of the spectrum for $\nu=2,3$ for strong dissipative coupling, $g=2\delta\omega=0.1$, and moderate reactive coupling, $J = 0.06$, at the transition regime between the PT broken and unbroken regimes, resulting in spectral flow involving large numbers of eigenvalues. Computed with $A=B=\omega_0 = 1, C=0.1$.
    \label{fig:fig8}}
\end{figure}

In the PT-broken phase, with strong reactive coupling, the long-time behavior changes significantly with the phase. Near $\phi_J = 0,\pi$ there is a single dominant eigenvalue, with a gap given by the dissipative coupling, $g$. The dynamics, on timescales $\gtrsim 1/g$, is thus synchronized, showing a single dominant frequency. We can interpret this in terms of the relationship between the collective dissipation, with jump operator $a_1-a_2$, and the eigenmodes of the reactive coupling, $a_1\pm a_2$. One eigenmode is orthogonal to the jump operator, and hence unaffected by the collective dissipation, while the other is parallel to it, and hence is maximally affected by it. The opposite situation arises near $\phi = \pm \pi/2$, where the two dominant eigenvalues have the same real part, suggesting the modes are equivalent with respect to the collective dissipation. The presence of these two eigenvalues, with identical real parts, implies the existence of two frequencies in the spectrum and, as in the purely reactive case, allows bistability. This is reminiscent of a synchronization blockade~\cite{kehrer2025}, where synchronization is suppressed leading to bistable phase-locking. 

As mentioned above, the behaviour of the two leading eigenvalues in the $\nu=1$ case implies the presence of an EP, where the two loops merge. Encircling this exceptional point leads to spectral flow where the two eigenvalues are exchanged as the phase turns one period. In open systems, spectral flow leads to unique behavior such as chiral state transfer \cite{milburn2015,schumer2022,sun2023,sun2024}, where eigenvectors can be exchanged after one cycle, but only when encircling the EP in a particular direction. In higher harmonics, shown in Fig.~\ref{fig:fig8} for $\nu=2,3$, we observe spectral flow involving a large number of eigenvalues. Similar behavior is observed for $\nu=1$ with eigenvalues with larger dissipation but for the parameters studied here the leading eigenvalues are always isolated from the bulk.

\subsection{Combined couplings: semiclassical approximation}

\label{sec:comb-coupl-semicl}
The spectral flows that occur with both types of coupling, and the exceptional points at $J\neq 0$, can be connected to corresponding phenomena in the semiclassical approximation. With the reactive coupling the linearized equations-of-motion in the semiclassical approximation, Eq. ~\eqref{eq:lineom}, become \begin{align}
    \frac{d}{dt}\mqty(\delta\expval{a_1} \\ \delta\expval{a_2}) = \frac{1}{2}\mqty(i(2\omega_0+\delta\omega)  & g-2iJe^{i\phi_J} \\ g-2iJe^{-i\phi_J} & i(2\omega_0-\delta \omega))\mqty(\delta\expval{ a_1} \\ \delta\expval{a_2}),\label{eq:lineomrc}
\end{align} giving exceptional points when \begin{align}\left(-g^2 + 4J^2 + \delta\omega^2\right)^2 + 16 g^2 J^2 \cos^2(\phi_J)=0.\end{align} For a generic coupling phase $\cos(\phi_J)\neq 0$, and the exceptional points occur only for $J=0$, at $|\delta \omega|=g$ as discussed previously. However, for coupling phases $\phi_J=\pm \pi/2$, $\cos(\phi_J)=0$, giving exceptional points when $g^2=4J^2+\delta\omega^2$. Thus the exceptional points lie along two lines forming the boundary of an Arnold tongue, unless $\phi_J=\pm \pi/2$, in which case they lie on the surface of a cone. However, while this approximation explains the presence of exceptional points, it fails to predict the spectral flow found in Fig.~\ref{fig:fig8}, which differs from that for $\nu=1$ and involves large numbers of eigenvalues.

\section{Conclusion }\label{sec:conclusion}

Coupled quantum self-sustained oscillators, like their classical
counterparts, show a rich variety of steady-state behaviors and
transitions related to synchronization. We have shown, using the
example of coupled lasers or polariton condensates, that this
gives rise to distinct forms for the slow modes of the quantum
Liouvillian, with transitions between the forms occurring at
exceptional points. These are singular points of the quantum
Liouvillian, at which the non-local structure of the dynamics changes,
and as such resemble the generalized exceptional points of classical
nonlinear systems~\cite{weis2025}. In contrast to the semiclassical
case, however, the singular dynamics does not occur on a single manifold
in parameter space; rather, there is a cascade of exceptional points,
corresponding to different orders of correlation functions and
different dynamical observables. Our results illustrate how
transitions in the long-time dynamics of a driven-dissipative quantum
system occur through singularities of the Liouvillian. Classical EPs
and generalized EPs lead to important phenomena~\cite{sun2024,weis2025} including chiral mode
switching, non-reciprocity, and anomalous dynamics, and our work
provides a basis for exploring similar effects in the quantum regime.

In addition, our work gives rise to the possibility of defining sharp
thresholds for synchronization, and related transitions, even in the
quantum regime, and identifying those points experimentally. In
contrast to typical synchronization measures in the steady-state,
which give smooth crossovers over a range of coupling
strengths~\cite{mari2013,solanki2026}, the EPs we find occur at
specific values of coupling for specific observables, which can be
mapped to experimental signatures. The $\nu=\pm1$ EPs, for example,
correspond to the coupling strength where the emission spectrum (the
Fourier transform of the first-order coherence) changes from one with
two peaks in frequency to one with one. We predict that higher-order
spectra show similar behavior, at slightly different couplings, but
over a limited range. While variations in the model will shift the
EPs, they are protected by a generalized PT symmetry, and so will not
be eliminated. However, going between different states does not imply
going through an EP: similarly to a critical point in thermodynamics, one has a
singularity which allows the states to change form, but is not
necessarily encountered on a given experimental path between
states. Nonetheless, available experimental techniques provide
sufficient tunability to access these EPs in
condensates~\cite{ohadi2016a,ohadi2018,chestnov2019,liang2026,berloff2017a}
and lasers~\cite{kreinberg2019,cao2019}, allowing them to be used to
diagnose synchronization and their consequences to be explored. It
would be interesting, also, to apply similar analyses to other systems
in which quantum synchronization and time-crystals are being studied,
including trapped ions, spins, and quantum
simulators~\cite{schmolke2026}.

\begin{acknowledgments}
We acknowledge funding from Taighde \'Eireann -- Research Ireland
(21-FF-P/10142) and helpful discussions with H. Ohadi.
\end{acknowledgments}

In order to meet institutional and research funder open access
requirements, any accepted manuscript arising shall be open access
under a Creative Commons Attribution (CC BY) reuse license with zero
embargo.

Code and data supporting this article are openly available~\cite{data}.

\bibliography{Quantumsynchronization2.bib,addmaterials.bib}

\begin{thebibliography}{74}%
\makeatletter
\providecommand \@ifxundefined [1]{%
 \@ifx{#1\undefined}
}%
\providecommand \@ifnum [1]{%
 \ifnum #1\expandafter \@firstoftwo
 \else \expandafter \@secondoftwo
 \fi
}%
\providecommand \@ifx [1]{%
 \ifx #1\expandafter \@firstoftwo
 \else \expandafter \@secondoftwo
 \fi
}%
\providecommand \natexlab [1]{#1}%
\providecommand \enquote  [1]{``#1''}%
\providecommand \bibnamefont  [1]{#1}%
\providecommand \bibfnamefont [1]{#1}%
\providecommand \citenamefont [1]{#1}%
\providecommand \href@noop [0]{\@secondoftwo}%
\providecommand \href [0]{\begingroup \@sanitize@url \@href}%
\providecommand \@href[1]{\@@startlink{#1}\@@href}%
\providecommand \@@href[1]{\endgroup#1\@@endlink}%
\providecommand \@sanitize@url [0]{\catcode `\\12\catcode `\$12\catcode
  `\&12\catcode `\#12\catcode `\^12\catcode `\_12\catcode `\%12\relax}%
\providecommand \@@startlink[1]{}%
\providecommand \@@endlink[0]{}%
\providecommand \url  [0]{\begingroup\@sanitize@url \@url }%
\providecommand \@url [1]{\endgroup\@href {#1}{\urlprefix }}%
\providecommand \urlprefix  [0]{URL }%
\providecommand \Eprint [0]{\href }%
\providecommand \doibase [0]{https://doi.org/}%
\providecommand \selectlanguage [0]{\@gobble}%
\providecommand \bibinfo  [0]{\@secondoftwo}%
\providecommand \bibfield  [0]{\@secondoftwo}%
\providecommand \translation [1]{[#1]}%
\providecommand \BibitemOpen [0]{}%
\providecommand \bibitemStop [0]{}%
\providecommand \bibitemNoStop [0]{.\EOS\space}%
\providecommand \EOS [0]{\spacefactor3000\relax}%
\providecommand \BibitemShut  [1]{\csname bibitem#1\endcsname}%
\let\auto@bib@innerbib\@empty
\bibitem [{\citenamefont {Pikovsky}\ \emph {et~al.}(2001)\citenamefont
  {Pikovsky}, \citenamefont {Rosenblum},\ and\ \citenamefont
  {Kurths}}]{pikovsky2001}%
  \BibitemOpen
  \bibfield  {author} {\bibinfo {author} {\bibfnamefont {A.}~\bibnamefont
  {Pikovsky}}, \bibinfo {author} {\bibfnamefont {M.}~\bibnamefont
  {Rosenblum}},\ and\ \bibinfo {author} {\bibfnamefont {J.}~\bibnamefont
  {Kurths}},\ }\href {https://doi.org/10.1017/CBO9780511755743} {\emph
  {\bibinfo {title} {Synchronization: {{A Universal Concept}} in {{Nonlinear
  Sciences}}}}},\ \bibinfo {edition} {1st}\ ed.\ (\bibinfo  {publisher}
  {Cambridge University Press},\ \bibinfo {year} {2001})\BibitemShut {NoStop}%
\bibitem [{\citenamefont {Balanov}(2009)}]{balanov2009}%
  \BibitemOpen
  \bibinfo {editor} {\bibfnamefont {A.}~\bibnamefont {Balanov}},\ ed.,\
  \href@noop {} {\emph {\bibinfo {title} {Synchronization: From Simple to
  Complex}}},\ Springer Complexity\ (\bibinfo  {publisher} {Springer},\
  \bibinfo {address} {Berlin},\ \bibinfo {year} {2009})\BibitemShut {NoStop}%
\bibitem [{\citenamefont {Moroney}\ and\ \citenamefont
  {Eastham}(2021)}]{moroney2021}%
  \BibitemOpen
  \bibfield  {author} {\bibinfo {author} {\bibfnamefont {J.~P.}\ \bibnamefont
  {Moroney}}\ and\ \bibinfo {author} {\bibfnamefont {P.~R.}\ \bibnamefont
  {Eastham}},\ }\bibfield  {title} {\bibinfo {title} {Synchronization in
  disordered oscillator lattices: {{Nonequilibrium}} phase transition for
  driven-dissipative bosons},\ }\href
  {https://doi.org/10.1103/PhysRevResearch.3.043092} {\bibfield  {journal}
  {\bibinfo  {journal} {Phys. Rev. Res.}\ }\textbf {\bibinfo {volume} {3}},\
  \bibinfo {pages} {043092} (\bibinfo {year} {2021})}\BibitemShut {NoStop}%
\bibitem [{\citenamefont {Moroney}\ and\ \citenamefont
  {Eastham}(2023)}]{moroney2023}%
  \BibitemOpen
  \bibfield  {author} {\bibinfo {author} {\bibfnamefont {J.~P.}\ \bibnamefont
  {Moroney}}\ and\ \bibinfo {author} {\bibfnamefont {P.~R.}\ \bibnamefont
  {Eastham}},\ }\bibfield  {title} {\bibinfo {title} {Synchronization and
  spacetime vortices in one-dimensional driven-dissipative condensates and
  coupled oscillator models},\ }\href
  {https://doi.org/10.1103/PhysRevB.108.195302} {\bibfield  {journal} {\bibinfo
   {journal} {Phys. Rev. B}\ }\textbf {\bibinfo {volume} {108}},\ \bibinfo
  {pages} {195302} (\bibinfo {year} {2023})}\BibitemShut {NoStop}%
\bibitem [{\citenamefont {Vladimirov}\ \emph {et~al.}(2003)\citenamefont
  {Vladimirov}, \citenamefont {Kozyreff},\ and\ \citenamefont
  {Mandel}}]{vladimirov2003}%
  \BibitemOpen
  \bibfield  {author} {\bibinfo {author} {\bibfnamefont {A.~G.}\ \bibnamefont
  {Vladimirov}}, \bibinfo {author} {\bibfnamefont {G.}~\bibnamefont
  {Kozyreff}},\ and\ \bibinfo {author} {\bibfnamefont {P.}~\bibnamefont
  {Mandel}},\ }\bibfield  {title} {\bibinfo {title} {Synchronization of weakly
  stable oscillators and semiconductor laser arrays},\ }\href
  {https://doi.org/10.1209/epl/i2003-00115-8} {\bibfield  {journal} {\bibinfo
  {journal} {Europhys. Lett.}\ }\textbf {\bibinfo {volume} {61}},\ \bibinfo
  {pages} {613} (\bibinfo {year} {2003})}\BibitemShut {NoStop}%
\bibitem [{\citenamefont {Pfl{\"u}ger}\ \emph {et~al.}(2023)\citenamefont
  {Pfl{\"u}ger}, \citenamefont {Brunner}, \citenamefont {Heuser}, \citenamefont
  {Lott}, \citenamefont {Reitzenstein},\ and\ \citenamefont
  {Fischer}}]{pfluger2023}%
  \BibitemOpen
  \bibfield  {author} {\bibinfo {author} {\bibfnamefont {M.}~\bibnamefont
  {Pfl{\"u}ger}}, \bibinfo {author} {\bibfnamefont {D.}~\bibnamefont
  {Brunner}}, \bibinfo {author} {\bibfnamefont {T.}~\bibnamefont {Heuser}},
  \bibinfo {author} {\bibfnamefont {J.~A.}\ \bibnamefont {Lott}}, \bibinfo
  {author} {\bibfnamefont {S.}~\bibnamefont {Reitzenstein}},\ and\ \bibinfo
  {author} {\bibfnamefont {I.}~\bibnamefont {Fischer}},\ }\bibfield  {title}
  {\bibinfo {title} {Injection locking and coupling the emitters of large
  {{VCSEL}} arrays via diffraction in an external cavity},\ }\href
  {https://doi.org/10.1364/OE.473449} {\bibfield  {journal} {\bibinfo
  {journal} {Opt. Express}\ }\textbf {\bibinfo {volume} {31}},\ \bibinfo
  {pages} {8704} (\bibinfo {year} {2023})}\BibitemShut {NoStop}%
\bibitem [{\citenamefont {Ohadi}\ \emph {et~al.}(2018)\citenamefont {Ohadi},
  \citenamefont {{del Valle-Inclan Redondo}}, \citenamefont {Ramsay},
  \citenamefont {Hatzopoulos}, \citenamefont {Liew}, \citenamefont {Eastham},
  \citenamefont {Savvidis},\ and\ \citenamefont {Baumberg}}]{ohadi2018}%
  \BibitemOpen
  \bibfield  {author} {\bibinfo {author} {\bibfnamefont {H.}~\bibnamefont
  {Ohadi}}, \bibinfo {author} {\bibfnamefont {Y.}~\bibnamefont {{del
  Valle-Inclan Redondo}}}, \bibinfo {author} {\bibfnamefont {A.~J.}\
  \bibnamefont {Ramsay}}, \bibinfo {author} {\bibfnamefont {Z.}~\bibnamefont
  {Hatzopoulos}}, \bibinfo {author} {\bibfnamefont {T.~C.~H.}\ \bibnamefont
  {Liew}}, \bibinfo {author} {\bibfnamefont {P.~R.}\ \bibnamefont {Eastham}},
  \bibinfo {author} {\bibfnamefont {P.~G.}\ \bibnamefont {Savvidis}},\ and\
  \bibinfo {author} {\bibfnamefont {J.~J.}\ \bibnamefont {Baumberg}},\
  }\bibfield  {title} {\bibinfo {title} {Synchronization crossover of polariton
  condensates in weakly disordered lattices},\ }\href
  {https://doi.org/10.1103/PhysRevB.97.195109} {\bibfield  {journal} {\bibinfo
  {journal} {Phys. Rev. B}\ }\textbf {\bibinfo {volume} {97}},\ \bibinfo
  {pages} {195109} (\bibinfo {year} {2018})}\BibitemShut {NoStop}%
\bibitem [{\citenamefont {Chestnov}\ \emph {et~al.}(2019)\citenamefont
  {Chestnov}, \citenamefont {Kavokin},\ and\ \citenamefont
  {Yulin}}]{chestnov2019}%
  \BibitemOpen
  \bibfield  {author} {\bibinfo {author} {\bibfnamefont {I.~Y.}\ \bibnamefont
  {Chestnov}}, \bibinfo {author} {\bibfnamefont {A.~V.}\ \bibnamefont
  {Kavokin}},\ and\ \bibinfo {author} {\bibfnamefont {A.~V.}\ \bibnamefont
  {Yulin}},\ }\bibfield  {title} {\bibinfo {title} {The optical control of
  phase locking of polariton condensates},\ }\href
  {https://doi.org/10.1088/1367-2630/ab4d03} {\bibfield  {journal} {\bibinfo
  {journal} {New J. Phys.}\ }\textbf {\bibinfo {volume} {21}},\ \bibinfo
  {pages} {113009} (\bibinfo {year} {2019})}\BibitemShut {NoStop}%
\bibitem [{\citenamefont {Wouters}(2008)}]{wouters2008}%
  \BibitemOpen
  \bibfield  {author} {\bibinfo {author} {\bibfnamefont {M.}~\bibnamefont
  {Wouters}},\ }\bibfield  {title} {\bibinfo {title} {Synchronized and
  desynchronized phases of coupled nonequilibrium exciton-polariton
  condensates},\ }\href {https://doi.org/10.1103/PhysRevB.77.121302} {\bibfield
   {journal} {\bibinfo  {journal} {Phys. Rev. B}\ }\textbf {\bibinfo {volume}
  {77}},\ \bibinfo {pages} {121302} (\bibinfo {year} {2008})}\BibitemShut
  {NoStop}%
\bibitem [{\citenamefont {Eastham}(2008)}]{eastham2008}%
  \BibitemOpen
  \bibfield  {author} {\bibinfo {author} {\bibfnamefont {P.~R.}\ \bibnamefont
  {Eastham}},\ }\bibfield  {title} {\bibinfo {title} {Mode-locking and
  mode-competition in a non-equilibrium solid-state condensate},\ }\href
  {https://doi.org/10.1103/PhysRevB.78.035319} {\bibfield  {journal} {\bibinfo
  {journal} {Phys. Rev. B}\ }\textbf {\bibinfo {volume} {78}},\ \bibinfo
  {pages} {035319} (\bibinfo {year} {2008})}\BibitemShut {NoStop}%
\bibitem [{\citenamefont {Wiesenfeld}\ \emph {et~al.}(1996)\citenamefont
  {Wiesenfeld}, \citenamefont {Colet},\ and\ \citenamefont
  {Strogatz}}]{wiesenfeld1996}%
  \BibitemOpen
  \bibfield  {author} {\bibinfo {author} {\bibfnamefont {K.}~\bibnamefont
  {Wiesenfeld}}, \bibinfo {author} {\bibfnamefont {P.}~\bibnamefont {Colet}},\
  and\ \bibinfo {author} {\bibfnamefont {S.~H.}\ \bibnamefont {Strogatz}},\
  }\bibfield  {title} {\bibinfo {title} {Synchronization {{Transitions}} in a
  {{Disordered Josephson Series Array}}},\ }\href
  {https://doi.org/10.1103/PhysRevLett.76.404} {\bibfield  {journal} {\bibinfo
  {journal} {Phys. Rev. Lett.}\ }\textbf {\bibinfo {volume} {76}},\ \bibinfo
  {pages} {404} (\bibinfo {year} {1996})}\BibitemShut {NoStop}%
\bibitem [{\citenamefont {Heiss}(2012)}]{heiss2012}%
  \BibitemOpen
  \bibfield  {author} {\bibinfo {author} {\bibfnamefont {W.~D.}\ \bibnamefont
  {Heiss}},\ }\bibfield  {title} {\bibinfo {title} {The physics of exceptional
  points},\ }\href {https://doi.org/10.1088/1751-8113/45/44/444016} {\bibfield
  {journal} {\bibinfo  {journal} {J. Phys. Math. Theor.}\ }\textbf {\bibinfo
  {volume} {45}},\ \bibinfo {pages} {444016} (\bibinfo {year}
  {2012})}\BibitemShut {NoStop}%
\bibitem [{\citenamefont {Weinreich}(1977)}]{weinreich1977}%
  \BibitemOpen
  \bibfield  {author} {\bibinfo {author} {\bibfnamefont {G.}~\bibnamefont
  {Weinreich}},\ }\bibfield  {title} {\bibinfo {title} {Coupled piano
  strings},\ }\href {https://doi.org/10.1121/1.381677} {\bibfield  {journal}
  {\bibinfo  {journal} {J. Acoust. Soc. Am.}\ }\textbf {\bibinfo {volume}
  {62}},\ \bibinfo {pages} {1474} (\bibinfo {year} {1977})}\BibitemShut
  {NoStop}%
\bibitem [{\citenamefont {Weis}\ \emph {et~al.}(2025)\citenamefont {Weis},
  \citenamefont {Fruchart}, \citenamefont {Hanai}, \citenamefont {Kawagoe},
  \citenamefont {Littlewood},\ and\ \citenamefont {Vitelli}}]{weis2025}%
  \BibitemOpen
  \bibfield  {author} {\bibinfo {author} {\bibfnamefont {C.}~\bibnamefont
  {Weis}}, \bibinfo {author} {\bibfnamefont {M.}~\bibnamefont {Fruchart}},
  \bibinfo {author} {\bibfnamefont {R.}~\bibnamefont {Hanai}}, \bibinfo
  {author} {\bibfnamefont {K.}~\bibnamefont {Kawagoe}}, \bibinfo {author}
  {\bibfnamefont {P.~B.}\ \bibnamefont {Littlewood}},\ and\ \bibinfo {author}
  {\bibfnamefont {V.}~\bibnamefont {Vitelli}},\ }\bibfield  {title} {\bibinfo
  {title} {Generalized exceptional points in nonlinear and stochastic
  dynamics},\ }\href {https://doi.org/10.1103/mnn4-b298} {\bibfield  {journal}
  {\bibinfo  {journal} {Phys. Rev. Res.}\ }\textbf {\bibinfo {volume} {7}},\
  \bibinfo {pages} {043157} (\bibinfo {year} {2025})}\BibitemShut {NoStop}%
\bibitem [{\citenamefont {Whittaker}\ and\ \citenamefont
  {Eastham}(2009)}]{whittaker2009}%
  \BibitemOpen
  \bibfield  {author} {\bibinfo {author} {\bibfnamefont {D.~M.}\ \bibnamefont
  {Whittaker}}\ and\ \bibinfo {author} {\bibfnamefont {P.~R.}\ \bibnamefont
  {Eastham}},\ }\bibfield  {title} {\bibinfo {title} {Coherence properties of
  the microcavity polariton condensate},\ }\href
  {https://doi.org/10.1209/0295-5075/87/27002} {\bibfield  {journal} {\bibinfo
  {journal} {EPL Europhys. Lett.}\ }\textbf {\bibinfo {volume} {87}},\ \bibinfo
  {pages} {27002} (\bibinfo {year} {2009})}\BibitemShut {NoStop}%
\bibitem [{\citenamefont {Kreinberg}\ \emph {et~al.}(2019)\citenamefont
  {Kreinberg}, \citenamefont {Porte}, \citenamefont {Schicke}, \citenamefont
  {Lingnau}, \citenamefont {Schneider}, \citenamefont {H{\"o}fling},
  \citenamefont {Kanter}, \citenamefont {L{\"u}dge},\ and\ \citenamefont
  {Reitzenstein}}]{kreinberg2019}%
  \BibitemOpen
  \bibfield  {author} {\bibinfo {author} {\bibfnamefont {S.}~\bibnamefont
  {Kreinberg}}, \bibinfo {author} {\bibfnamefont {X.}~\bibnamefont {Porte}},
  \bibinfo {author} {\bibfnamefont {D.}~\bibnamefont {Schicke}}, \bibinfo
  {author} {\bibfnamefont {B.}~\bibnamefont {Lingnau}}, \bibinfo {author}
  {\bibfnamefont {C.}~\bibnamefont {Schneider}}, \bibinfo {author}
  {\bibfnamefont {S.}~\bibnamefont {H{\"o}fling}}, \bibinfo {author}
  {\bibfnamefont {I.}~\bibnamefont {Kanter}}, \bibinfo {author} {\bibfnamefont
  {K.}~\bibnamefont {L{\"u}dge}},\ and\ \bibinfo {author} {\bibfnamefont
  {S.}~\bibnamefont {Reitzenstein}},\ }\bibfield  {title} {\bibinfo {title}
  {Mutual coupling and synchronization of optically coupled quantum-dot
  micropillar lasers at ultra-low light levels},\ }\href
  {https://doi.org/10.1038/s41467-019-09559-2} {\bibfield  {journal} {\bibinfo
  {journal} {Nat. Commun.}\ }\textbf {\bibinfo {volume} {10}},\ \bibinfo
  {pages} {1539} (\bibinfo {year} {2019})}\BibitemShut {NoStop}%
\bibitem [{\citenamefont {Scully}\ and\ \citenamefont
  {Zubairy}(1997)}]{scully1997}%
  \BibitemOpen
  \bibfield  {author} {\bibinfo {author} {\bibfnamefont {M.~O.}\ \bibnamefont
  {Scully}}\ and\ \bibinfo {author} {\bibfnamefont {M.~S.}\ \bibnamefont
  {Zubairy}},\ }\href {https://doi.org/10.1017/CBO9780511813993} {\emph
  {\bibinfo {title} {Quantum {{Optics}}}}},\ \bibinfo {edition} {1st}\ ed.\
  (\bibinfo  {publisher} {Cambridge University Press},\ \bibinfo {year}
  {1997})\BibitemShut {NoStop}%
\bibitem [{\citenamefont {Ding}\ \emph {et~al.}(2019)\citenamefont {Ding},
  \citenamefont {Belykh}, \citenamefont {Marandi},\ and\ \citenamefont
  {Miri}}]{ding2019}%
  \BibitemOpen
  \bibfield  {author} {\bibinfo {author} {\bibfnamefont {J.}~\bibnamefont
  {Ding}}, \bibinfo {author} {\bibfnamefont {I.}~\bibnamefont {Belykh}},
  \bibinfo {author} {\bibfnamefont {A.}~\bibnamefont {Marandi}},\ and\ \bibinfo
  {author} {\bibfnamefont {M.-A.}\ \bibnamefont {Miri}},\ }\bibfield  {title}
  {\bibinfo {title} {Dispersive versus {{Dissipative Coupling}} for {{Frequency
  Synchronization}} in {{Lasers}}},\ }\href
  {https://doi.org/10.1103/PhysRevApplied.12.054039} {\bibfield  {journal}
  {\bibinfo  {journal} {Phys. Rev. Appl.}\ }\textbf {\bibinfo {volume} {12}},\
  \bibinfo {pages} {054039} (\bibinfo {year} {2019})}\BibitemShut {NoStop}%
\bibitem [{\citenamefont {Siegman}(1986)}]{siegman1986}%
  \BibitemOpen
  \bibfield  {author} {\bibinfo {author} {\bibfnamefont {A.~E.}\ \bibnamefont
  {Siegman}},\ }\href@noop {} {\emph {\bibinfo {title} {Lasers}}}\ (\bibinfo
  {publisher} {University Science Books},\ \bibinfo {address} {Mill Valley,
  California},\ \bibinfo {year} {1986})\BibitemShut {NoStop}%
\bibitem [{\citenamefont {Minganti}\ \emph {et~al.}(2018)\citenamefont
  {Minganti}, \citenamefont {Biella}, \citenamefont {Bartolo},\ and\
  \citenamefont {Ciuti}}]{minganti2018a}%
  \BibitemOpen
  \bibfield  {author} {\bibinfo {author} {\bibfnamefont {F.}~\bibnamefont
  {Minganti}}, \bibinfo {author} {\bibfnamefont {A.}~\bibnamefont {Biella}},
  \bibinfo {author} {\bibfnamefont {N.}~\bibnamefont {Bartolo}},\ and\ \bibinfo
  {author} {\bibfnamefont {C.}~\bibnamefont {Ciuti}},\ }\bibfield  {title}
  {\bibinfo {title} {Spectral theory of {{Liouvillians}} for dissipative phase
  transitions},\ }\href {https://doi.org/10.1103/PhysRevA.98.042118} {\bibfield
   {journal} {\bibinfo  {journal} {Phys. Rev. A}\ }\textbf {\bibinfo {volume}
  {98}},\ \bibinfo {pages} {042118} (\bibinfo {year} {2018})}\BibitemShut
  {NoStop}%
\bibitem [{\citenamefont {Cabot}\ \emph {et~al.}(2024)\citenamefont {Cabot},
  \citenamefont {Giorgi},\ and\ \citenamefont {Zambrini}}]{cabot2024}%
  \BibitemOpen
  \bibfield  {author} {\bibinfo {author} {\bibfnamefont {A.}~\bibnamefont
  {Cabot}}, \bibinfo {author} {\bibfnamefont {G.~L.}\ \bibnamefont {Giorgi}},\
  and\ \bibinfo {author} {\bibfnamefont {R.}~\bibnamefont {Zambrini}},\
  }\bibfield  {title} {\bibinfo {title} {Nonequilibrium {{Transition}} between
  {{Dissipative Time Crystals}}},\ }\href
  {https://doi.org/10.1103/PRXQuantum.5.030325} {\bibfield  {journal} {\bibinfo
   {journal} {PRX Quantum}\ }\textbf {\bibinfo {volume} {5}},\ \bibinfo {pages}
  {030325} (\bibinfo {year} {2024})}\BibitemShut {NoStop}%
\bibitem [{\citenamefont {Dutta}\ \emph {et~al.}(2025)\citenamefont {Dutta},
  \citenamefont {Zhang},\ and\ \citenamefont {Haque}}]{dutta2025}%
  \BibitemOpen
  \bibfield  {author} {\bibinfo {author} {\bibfnamefont {S.}~\bibnamefont
  {Dutta}}, \bibinfo {author} {\bibfnamefont {S.}~\bibnamefont {Zhang}},\ and\
  \bibinfo {author} {\bibfnamefont {M.}~\bibnamefont {Haque}},\ }\bibfield
  {title} {\bibinfo {title} {Quantum {{Origin}} of {{Limit Cycles}}, {{Fixed
  Points}}, and {{Critical Slowing Down}}},\ }\href
  {https://doi.org/10.1103/PhysRevLett.134.050407} {\bibfield  {journal}
  {\bibinfo  {journal} {Phys. Rev. Lett.}\ }\textbf {\bibinfo {volume} {134}},\
  \bibinfo {pages} {050407} (\bibinfo {year} {2025})}\BibitemShut {NoStop}%
\bibitem [{\citenamefont {Alaeian}\ \emph {et~al.}(2021)\citenamefont
  {Alaeian}, \citenamefont {Giedke}, \citenamefont {Carusotto}, \citenamefont
  {L{\"o}w},\ and\ \citenamefont {Pfau}}]{alaeian2021}%
  \BibitemOpen
  \bibfield  {author} {\bibinfo {author} {\bibfnamefont {H.}~\bibnamefont
  {Alaeian}}, \bibinfo {author} {\bibfnamefont {G.}~\bibnamefont {Giedke}},
  \bibinfo {author} {\bibfnamefont {I.}~\bibnamefont {Carusotto}}, \bibinfo
  {author} {\bibfnamefont {R.}~\bibnamefont {L{\"o}w}},\ and\ \bibinfo {author}
  {\bibfnamefont {T.}~\bibnamefont {Pfau}},\ }\bibfield  {title} {\bibinfo
  {title} {Limit {{Cycle Phase}} and {{Goldstone Mode}} in {{Driven Dissipative
  Systems}}},\ }\href {https://doi.org/10.1103/PhysRevA.103.013712} {\bibfield
  {journal} {\bibinfo  {journal} {Phys. Rev. A}\ }\textbf {\bibinfo {volume}
  {103}},\ \bibinfo {pages} {013712} (\bibinfo {year} {2021})}\BibitemShut
  {NoStop}%
\bibitem [{\citenamefont {Schmolke}\ and\ \citenamefont
  {Lutz}(2026)}]{schmolke2026}%
  \BibitemOpen
  \bibfield  {author} {\bibinfo {author} {\bibfnamefont {F.}~\bibnamefont
  {Schmolke}}\ and\ \bibinfo {author} {\bibfnamefont {E.}~\bibnamefont
  {Lutz}},\ }\bibfield  {title} {\bibinfo {title} {Synchronization in the
  quantum regime},\ }\bibfield  {journal} {\bibinfo  {journal}
  {arXiv:2606.21226}\ }\href {https://doi.org/10.48550/arXiv.2606.21226}
  {10.48550/arXiv.2606.21226} (\bibinfo {year} {2026})\BibitemShut {NoStop}%
\bibitem [{\citenamefont {Solanki}\ \emph {et~al.}(2026)\citenamefont
  {Solanki}, \citenamefont {Cabot}, \citenamefont {Iemini}, \citenamefont
  {Carollo}, \citenamefont {Krishna}, \citenamefont {Ibrahim}, \citenamefont
  {Hajdu{\v s}ek}, \citenamefont {Lesanovsky}, \citenamefont {Fazio},
  \citenamefont {Zambrini},\ and\ \citenamefont {Vinjanampathy}}]{solanki2026}%
  \BibitemOpen
  \bibfield  {author} {\bibinfo {author} {\bibfnamefont {P.}~\bibnamefont
  {Solanki}}, \bibinfo {author} {\bibfnamefont {A.}~\bibnamefont {Cabot}},
  \bibinfo {author} {\bibfnamefont {F.}~\bibnamefont {Iemini}}, \bibinfo
  {author} {\bibfnamefont {F.}~\bibnamefont {Carollo}}, \bibinfo {author}
  {\bibfnamefont {M.}~\bibnamefont {Krishna}}, \bibinfo {author} {\bibfnamefont
  {Y.}~\bibnamefont {Ibrahim}}, \bibinfo {author} {\bibfnamefont
  {M.}~\bibnamefont {Hajdu{\v s}ek}}, \bibinfo {author} {\bibfnamefont
  {I.}~\bibnamefont {Lesanovsky}}, \bibinfo {author} {\bibfnamefont
  {R.}~\bibnamefont {Fazio}}, \bibinfo {author} {\bibfnamefont
  {R.}~\bibnamefont {Zambrini}},\ and\ \bibinfo {author} {\bibfnamefont
  {S.}~\bibnamefont {Vinjanampathy}},\ }\bibfield  {title} {\bibinfo {title}
  {Quantum {{Synchronization}}},\ }\bibfield  {journal} {\bibinfo  {journal}
  {arXiv:2607.19328}\ }\href {https://doi.org/10.48550/arXiv.2607.19328}
  {10.48550/arXiv.2607.19328} (\bibinfo {year} {2026})\BibitemShut {NoStop}%
\bibitem [{\citenamefont {Lee}\ and\ \citenamefont
  {Sadeghpour}(2013)}]{lee2013}%
  \BibitemOpen
  \bibfield  {author} {\bibinfo {author} {\bibfnamefont {T.~E.}\ \bibnamefont
  {Lee}}\ and\ \bibinfo {author} {\bibfnamefont {H.~R.}\ \bibnamefont
  {Sadeghpour}},\ }\bibfield  {title} {\bibinfo {title} {Quantum
  {{Synchronization}} of {{Quantum}} van der {{Pol Oscillators}} with {{Trapped
  Ions}}},\ }\href {https://doi.org/10.1103/PhysRevLett.111.234101} {\bibfield
  {journal} {\bibinfo  {journal} {Phys. Rev. Lett.}\ }\textbf {\bibinfo
  {volume} {111}},\ \bibinfo {pages} {234101} (\bibinfo {year}
  {2013})}\BibitemShut {NoStop}%
\bibitem [{\citenamefont {Ben~Arosh}\ \emph {et~al.}(2021)\citenamefont
  {Ben~Arosh}, \citenamefont {Cross},\ and\ \citenamefont
  {Lifshitz}}]{benarosh2021}%
  \BibitemOpen
  \bibfield  {author} {\bibinfo {author} {\bibfnamefont {L.}~\bibnamefont
  {Ben~Arosh}}, \bibinfo {author} {\bibfnamefont {M.~C.}\ \bibnamefont
  {Cross}},\ and\ \bibinfo {author} {\bibfnamefont {R.}~\bibnamefont
  {Lifshitz}},\ }\bibfield  {title} {\bibinfo {title} {Quantum limit cycles and
  the {{Rayleigh}} and van der {{Pol}} oscillators},\ }\href
  {https://doi.org/10.1103/PhysRevResearch.3.013130} {\bibfield  {journal}
  {\bibinfo  {journal} {Phys. Rev. Res.}\ }\textbf {\bibinfo {volume} {3}},\
  \bibinfo {pages} {013130} (\bibinfo {year} {2021})}\BibitemShut {NoStop}%
\bibitem [{\citenamefont {Chia}\ \emph {et~al.}(2020)\citenamefont {Chia},
  \citenamefont {Kwek},\ and\ \citenamefont {Noh}}]{chia2020}%
  \BibitemOpen
  \bibfield  {author} {\bibinfo {author} {\bibfnamefont {A.}~\bibnamefont
  {Chia}}, \bibinfo {author} {\bibfnamefont {L.~C.}\ \bibnamefont {Kwek}},\
  and\ \bibinfo {author} {\bibfnamefont {C.}~\bibnamefont {Noh}},\ }\bibfield
  {title} {\bibinfo {title} {Relaxation oscillations and frequency entrainment
  in quantum mechanics},\ }\href {https://doi.org/10.1103/PhysRevE.102.042213}
  {\bibfield  {journal} {\bibinfo  {journal} {Phys. Rev. E}\ }\textbf {\bibinfo
  {volume} {102}},\ \bibinfo {pages} {042213} (\bibinfo {year}
  {2020})}\BibitemShut {NoStop}%
\bibitem [{\citenamefont {Mari}\ \emph {et~al.}(2013)\citenamefont {Mari},
  \citenamefont {Farace}, \citenamefont {Didier}, \citenamefont {Giovannetti},\
  and\ \citenamefont {Fazio}}]{mari2013}%
  \BibitemOpen
  \bibfield  {author} {\bibinfo {author} {\bibfnamefont {A.}~\bibnamefont
  {Mari}}, \bibinfo {author} {\bibfnamefont {A.}~\bibnamefont {Farace}},
  \bibinfo {author} {\bibfnamefont {N.}~\bibnamefont {Didier}}, \bibinfo
  {author} {\bibfnamefont {V.}~\bibnamefont {Giovannetti}},\ and\ \bibinfo
  {author} {\bibfnamefont {R.}~\bibnamefont {Fazio}},\ }\bibfield  {title}
  {\bibinfo {title} {Measures of {{Quantum Synchronization}} in {{Continuous
  Variable Systems}}},\ }\href {https://doi.org/10.1103/PhysRevLett.111.103605}
  {\bibfield  {journal} {\bibinfo  {journal} {Phys. Rev. Lett.}\ }\textbf
  {\bibinfo {volume} {111}},\ \bibinfo {pages} {103605} (\bibinfo {year}
  {2013})}\BibitemShut {NoStop}%
\bibitem [{\citenamefont {Walter}\ \emph {et~al.}(2014)\citenamefont {Walter},
  \citenamefont {Nunnenkamp},\ and\ \citenamefont {Bruder}}]{walter2014}%
  \BibitemOpen
  \bibfield  {author} {\bibinfo {author} {\bibfnamefont {S.}~\bibnamefont
  {Walter}}, \bibinfo {author} {\bibfnamefont {A.}~\bibnamefont {Nunnenkamp}},\
  and\ \bibinfo {author} {\bibfnamefont {C.}~\bibnamefont {Bruder}},\
  }\bibfield  {title} {\bibinfo {title} {Quantum synchronization of a driven
  self-sustained oscillator},\ }\href
  {https://doi.org/10.1103/PhysRevLett.112.094102} {\bibfield  {journal}
  {\bibinfo  {journal} {Phys. Rev. Lett.}\ }\textbf {\bibinfo {volume} {112}},\
  \bibinfo {pages} {094102} (\bibinfo {year} {2014})}\BibitemShut {NoStop}%
\bibitem [{\citenamefont {Walter}\ \emph {et~al.}(2015)\citenamefont {Walter},
  \citenamefont {Nunnenkamp},\ and\ \citenamefont {Bruder}}]{walter2015}%
  \BibitemOpen
  \bibfield  {author} {\bibinfo {author} {\bibfnamefont {S.}~\bibnamefont
  {Walter}}, \bibinfo {author} {\bibfnamefont {A.}~\bibnamefont {Nunnenkamp}},\
  and\ \bibinfo {author} {\bibfnamefont {C.}~\bibnamefont {Bruder}},\
  }\bibfield  {title} {\bibinfo {title} {Quantum synchronization of two {{Van}}
  der {{Pol}} oscillators},\ }\href {https://doi.org/10.1002/andp.201400144}
  {\bibfield  {journal} {\bibinfo  {journal} {Ann. Phys.}\ }\textbf {\bibinfo
  {volume} {527}},\ \bibinfo {pages} {131} (\bibinfo {year}
  {2015})}\BibitemShut {NoStop}%
\bibitem [{\citenamefont {Koppenh{\"o}fer}\ and\ \citenamefont
  {Roulet}(2019)}]{koppenhofer2019}%
  \BibitemOpen
  \bibfield  {author} {\bibinfo {author} {\bibfnamefont {M.}~\bibnamefont
  {Koppenh{\"o}fer}}\ and\ \bibinfo {author} {\bibfnamefont {A.}~\bibnamefont
  {Roulet}},\ }\bibfield  {title} {\bibinfo {title} {Optimal synchronization
  deep in the quantum regime: Resource and fundamental limit},\ }\href
  {https://doi.org/10.1103/PhysRevA.99.043804} {\bibfield  {journal} {\bibinfo
  {journal} {Phys. Rev. A}\ }\textbf {\bibinfo {volume} {99}},\ \bibinfo
  {pages} {043804} (\bibinfo {year} {2019})}\BibitemShut {NoStop}%
\bibitem [{\citenamefont {Jaseem}\ \emph {et~al.}(2020)\citenamefont {Jaseem},
  \citenamefont {Hajdu{\v s}ek}, \citenamefont {Vedral}, \citenamefont {Fazio},
  \citenamefont {Kwek},\ and\ \citenamefont {Vinjanampathy}}]{jaseem2020}%
  \BibitemOpen
  \bibfield  {author} {\bibinfo {author} {\bibfnamefont {N.}~\bibnamefont
  {Jaseem}}, \bibinfo {author} {\bibfnamefont {M.}~\bibnamefont {Hajdu{\v
  s}ek}}, \bibinfo {author} {\bibfnamefont {V.}~\bibnamefont {Vedral}},
  \bibinfo {author} {\bibfnamefont {R.}~\bibnamefont {Fazio}}, \bibinfo
  {author} {\bibfnamefont {L.-C.}\ \bibnamefont {Kwek}},\ and\ \bibinfo
  {author} {\bibfnamefont {S.}~\bibnamefont {Vinjanampathy}},\ }\bibfield
  {title} {\bibinfo {title} {Quantum synchronization in nanoscale heat
  engines},\ }\href {https://doi.org/10.1103/PhysRevE.101.020201} {\bibfield
  {journal} {\bibinfo  {journal} {Phys. Rev. E}\ }\textbf {\bibinfo {volume}
  {101}},\ \bibinfo {pages} {020201} (\bibinfo {year} {2020})}\BibitemShut
  {NoStop}%
\bibitem [{\citenamefont {{Manju}}\ \emph {et~al.}(2023)\citenamefont
  {{Manju}}, \citenamefont {Dasgupta},\ and\ \citenamefont
  {Biswas}}]{manju2023}%
  \BibitemOpen
  \bibfield  {author} {\bibinfo {author} {\bibnamefont {{Manju}}}, \bibinfo
  {author} {\bibfnamefont {S.}~\bibnamefont {Dasgupta}},\ and\ \bibinfo
  {author} {\bibfnamefont {A.}~\bibnamefont {Biswas}},\ }\bibfield  {title}
  {\bibinfo {title} {Entanglement boosts quantum synchronization between two
  oscillators in an optomechanical setup},\ }\href
  {https://doi.org/10.1016/j.physleta.2023.129039} {\bibfield  {journal}
  {\bibinfo  {journal} {Phys. Lett. A}\ }\textbf {\bibinfo {volume} {482}},\
  \bibinfo {pages} {129039} (\bibinfo {year} {2023})}\BibitemShut {NoStop}%
\bibitem [{\citenamefont {Hong}\ \emph {et~al.}(2024)\citenamefont {Hong},
  \citenamefont {Wu}, \citenamefont {Peng}, \citenamefont {Qian}, \citenamefont
  {Hu},\ and\ \citenamefont {Wang}}]{hong2024}%
  \BibitemOpen
  \bibfield  {author} {\bibinfo {author} {\bibfnamefont {Q.}~\bibnamefont
  {Hong}}, \bibinfo {author} {\bibfnamefont {W.-X.}\ \bibnamefont {Wu}},
  \bibinfo {author} {\bibfnamefont {Y.-P.}\ \bibnamefont {Peng}}, \bibinfo
  {author} {\bibfnamefont {J.}~\bibnamefont {Qian}}, \bibinfo {author}
  {\bibfnamefont {C.-M.}\ \bibnamefont {Hu}},\ and\ \bibinfo {author}
  {\bibfnamefont {Y.-P.}\ \bibnamefont {Wang}},\ }\bibfield  {title} {\bibinfo
  {title} {Synchronization and asynchronization in non-{{Hermitian}} systems},\
  }\href {https://doi.org/10.1103/PhysRevA.110.052201} {\bibfield  {journal}
  {\bibinfo  {journal} {Phys. Rev. A}\ }\textbf {\bibinfo {volume} {110}},\
  \bibinfo {pages} {052201} (\bibinfo {year} {2024})}\BibitemShut {NoStop}%
\bibitem [{\citenamefont {Kehrer}\ and\ \citenamefont
  {Bruder}(2025)}]{kehrer2025}%
  \BibitemOpen
  \bibfield  {author} {\bibinfo {author} {\bibfnamefont {T.}~\bibnamefont
  {Kehrer}}\ and\ \bibinfo {author} {\bibfnamefont {C.}~\bibnamefont
  {Bruder}},\ }\bibfield  {title} {\bibinfo {title} {Quantum synchronization
  blockade induced by nonreciprocal coupling},\ }\href
  {https://doi.org/10.1103/9y9w-lw92} {\bibfield  {journal} {\bibinfo
  {journal} {Phys. Rev. A}\ }\textbf {\bibinfo {volume} {112}},\ \bibinfo
  {pages} {012223} (\bibinfo {year} {2025})}\BibitemShut {NoStop}%
\bibitem [{\citenamefont {Lai}\ \emph {et~al.}(2025)\citenamefont {Lai},
  \citenamefont {Miranowicz},\ and\ \citenamefont {Nori}}]{lai2025}%
  \BibitemOpen
  \bibfield  {author} {\bibinfo {author} {\bibfnamefont {D.-G.}\ \bibnamefont
  {Lai}}, \bibinfo {author} {\bibfnamefont {A.}~\bibnamefont {Miranowicz}},\
  and\ \bibinfo {author} {\bibfnamefont {F.}~\bibnamefont {Nori}},\ }\bibfield
  {title} {\bibinfo {title} {Nonreciprocal quantum synchronization},\ }\href
  {https://doi.org/10.1038/s41467-025-63408-z} {\bibfield  {journal} {\bibinfo
  {journal} {Nat. Commun.}\ }\textbf {\bibinfo {volume} {16}},\ \bibinfo
  {pages} {8491} (\bibinfo {year} {2025})}\BibitemShut {NoStop}%
\bibitem [{\citenamefont {Nadolny}\ and\ \citenamefont
  {Bruder}(2026)}]{nadolny2026}%
  \BibitemOpen
  \bibfield  {author} {\bibinfo {author} {\bibfnamefont {T.}~\bibnamefont
  {Nadolny}}\ and\ \bibinfo {author} {\bibfnamefont {C.}~\bibnamefont
  {Bruder}},\ }\bibfield  {title} {\bibinfo {title} {Quantum limit cycles and
  synchronization from a measurement perspective},\ }\href
  {https://doi.org/10.1103/c5wj-4s73} {\bibfield  {journal} {\bibinfo
  {journal} {Phys. Rev. Res.}\ }\textbf {\bibinfo {volume} {8}},\ \bibinfo
  {pages} {023050} (\bibinfo {year} {2026})}\BibitemShut {NoStop}%
\bibitem [{\citenamefont {Roulet}\ and\ \citenamefont
  {Bruder}(2018{\natexlab{a}})}]{roulet2018a}%
  \BibitemOpen
  \bibfield  {author} {\bibinfo {author} {\bibfnamefont {A.}~\bibnamefont
  {Roulet}}\ and\ \bibinfo {author} {\bibfnamefont {C.}~\bibnamefont
  {Bruder}},\ }\bibfield  {title} {\bibinfo {title} {Synchronizing the
  {{Smallest Possible System}}},\ }\href
  {https://doi.org/10.1103/PhysRevLett.121.053601} {\bibfield  {journal}
  {\bibinfo  {journal} {Phys. Rev. Lett.}\ }\textbf {\bibinfo {volume} {121}},\
  \bibinfo {pages} {053601} (\bibinfo {year} {2018}{\natexlab{a}})}\BibitemShut
  {NoStop}%
\bibitem [{\citenamefont {Roulet}\ and\ \citenamefont
  {Bruder}(2018{\natexlab{b}})}]{roulet2018b}%
  \BibitemOpen
  \bibfield  {author} {\bibinfo {author} {\bibfnamefont {A.}~\bibnamefont
  {Roulet}}\ and\ \bibinfo {author} {\bibfnamefont {C.}~\bibnamefont
  {Bruder}},\ }\bibfield  {title} {\bibinfo {title} {Quantum synchronization
  and entanglement generation},\ }\href
  {https://doi.org/10.1103/PhysRevLett.121.063601} {\bibfield  {journal}
  {\bibinfo  {journal} {Phys. Rev. Lett.}\ }\textbf {\bibinfo {volume} {121}},\
  \bibinfo {pages} {063601} (\bibinfo {year} {2018}{\natexlab{b}})}\BibitemShut
  {NoStop}%
\bibitem [{\citenamefont {{Parra-L{\'o}pez}}\ and\ \citenamefont
  {Bergli}(2020)}]{parra-lopez2020}%
  \BibitemOpen
  \bibfield  {author} {\bibinfo {author} {\bibfnamefont {{\'A}.}~\bibnamefont
  {{Parra-L{\'o}pez}}}\ and\ \bibinfo {author} {\bibfnamefont {J.}~\bibnamefont
  {Bergli}},\ }\bibfield  {title} {\bibinfo {title} {Synchronization in
  two-level quantum systems},\ }\href
  {https://doi.org/10.1103/PhysRevA.101.062104} {\bibfield  {journal} {\bibinfo
   {journal} {Phys. Rev. A}\ }\textbf {\bibinfo {volume} {101}},\ \bibinfo
  {pages} {062104} (\bibinfo {year} {2020})}\BibitemShut {NoStop}%
\bibitem [{\citenamefont {Cabot}\ \emph {et~al.}(2021)\citenamefont {Cabot},
  \citenamefont {Luca~Giorgi},\ and\ \citenamefont {Zambrini}}]{cabot2021}%
  \BibitemOpen
  \bibfield  {author} {\bibinfo {author} {\bibfnamefont {A.}~\bibnamefont
  {Cabot}}, \bibinfo {author} {\bibfnamefont {G.}~\bibnamefont {Luca~Giorgi}},\
  and\ \bibinfo {author} {\bibfnamefont {R.}~\bibnamefont {Zambrini}},\
  }\bibfield  {title} {\bibinfo {title} {Synchronization and coalescence in a
  dissipative two-qubit system},\ }\href
  {https://doi.org/10.1098/rspa.2020.0850} {\bibfield  {journal} {\bibinfo
  {journal} {Proc. R. Soc. Math. Phys. Eng. Sci.}\ }\textbf {\bibinfo {volume}
  {477}},\ \bibinfo {pages} {20200850} (\bibinfo {year} {2021})}\BibitemShut
  {NoStop}%
\bibitem [{\citenamefont {Zhang}\ \emph {et~al.}(2023)\citenamefont {Zhang},
  \citenamefont {Wang}, \citenamefont {Wang}, \citenamefont {Zhang},
  \citenamefont {Wu}, \citenamefont {Jie},\ and\ \citenamefont
  {Lu}}]{zhang2023}%
  \BibitemOpen
  \bibfield  {author} {\bibinfo {author} {\bibfnamefont {L.}~\bibnamefont
  {Zhang}}, \bibinfo {author} {\bibfnamefont {Z.}~\bibnamefont {Wang}},
  \bibinfo {author} {\bibfnamefont {Y.}~\bibnamefont {Wang}}, \bibinfo {author}
  {\bibfnamefont {J.}~\bibnamefont {Zhang}}, \bibinfo {author} {\bibfnamefont
  {Z.}~\bibnamefont {Wu}}, \bibinfo {author} {\bibfnamefont {J.}~\bibnamefont
  {Jie}},\ and\ \bibinfo {author} {\bibfnamefont {Y.}~\bibnamefont {Lu}},\
  }\bibfield  {title} {\bibinfo {title} {Quantum synchronization of a single
  trapped-ion qubit},\ }\href
  {https://doi.org/10.1103/PhysRevResearch.5.033209} {\bibfield  {journal}
  {\bibinfo  {journal} {Phys. Rev. Res.}\ }\textbf {\bibinfo {volume} {5}},\
  \bibinfo {pages} {033209} (\bibinfo {year} {2023})}\BibitemShut {NoStop}%
\bibitem [{\citenamefont {Zhou}\ \emph {et~al.}(2023)\citenamefont {Zhou},
  \citenamefont {Zou},\ and\ \citenamefont {Shao}}]{zhou2023}%
  \BibitemOpen
  \bibfield  {author} {\bibinfo {author} {\bibfnamefont {K.-J.}\ \bibnamefont
  {Zhou}}, \bibinfo {author} {\bibfnamefont {J.}~\bibnamefont {Zou}},\ and\
  \bibinfo {author} {\bibfnamefont {B.}~\bibnamefont {Shao}},\ }\bibfield
  {title} {\bibinfo {title} {Dynamical transition between synchronization and
  antisynchronization with exceptional points},\ }\href
  {https://doi.org/10.1103/PhysRevA.108.042206} {\bibfield  {journal} {\bibinfo
   {journal} {Phys. Rev. A}\ }\textbf {\bibinfo {volume} {108}},\ \bibinfo
  {pages} {042206} (\bibinfo {year} {2023})}\BibitemShut {NoStop}%
\bibitem [{\citenamefont {Belyansky}\ \emph {et~al.}(2025)\citenamefont
  {Belyansky}, \citenamefont {Weis}, \citenamefont {Hanai}, \citenamefont
  {Littlewood},\ and\ \citenamefont {Clerk}}]{belyansky2025}%
  \BibitemOpen
  \bibfield  {author} {\bibinfo {author} {\bibfnamefont {R.}~\bibnamefont
  {Belyansky}}, \bibinfo {author} {\bibfnamefont {C.}~\bibnamefont {Weis}},
  \bibinfo {author} {\bibfnamefont {R.}~\bibnamefont {Hanai}}, \bibinfo
  {author} {\bibfnamefont {P.~B.}\ \bibnamefont {Littlewood}},\ and\ \bibinfo
  {author} {\bibfnamefont {A.~A.}\ \bibnamefont {Clerk}},\ }\bibfield  {title}
  {\bibinfo {title} {Phase {{Transitions}} in {{Nonreciprocal
  Driven-Dissipative Condensates}}},\ }\href
  {https://doi.org/10.1103/gphr-d1bc} {\bibfield  {journal} {\bibinfo
  {journal} {Phys. Rev. Lett.}\ }\textbf {\bibinfo {volume} {135}},\ \bibinfo
  {pages} {123401} (\bibinfo {year} {2025})}\BibitemShut {NoStop}%
\bibitem [{\citenamefont {Eastham}\ \emph {et~al.}(2003)\citenamefont
  {Eastham}, \citenamefont {Szymanska},\ and\ \citenamefont
  {Littlewood}}]{eastham2003}%
  \BibitemOpen
  \bibfield  {author} {\bibinfo {author} {\bibfnamefont {P.~R.}\ \bibnamefont
  {Eastham}}, \bibinfo {author} {\bibfnamefont {M.~H.}\ \bibnamefont
  {Szymanska}},\ and\ \bibinfo {author} {\bibfnamefont {P.~B.}\ \bibnamefont
  {Littlewood}},\ }\bibfield  {title} {\bibinfo {title} {Phase-locking in
  quantum and classical oscillators: Polariton condensates, lasers, and arrays
  of {{Josephson}} junctions},\ }\href
  {https://doi.org/10.1016/S0038-1098(03)00338-7} {\bibfield  {journal}
  {\bibinfo  {journal} {Solid State Commun.}\ }\textbf {\bibinfo {volume}
  {127}},\ \bibinfo {pages} {117} (\bibinfo {year} {2003})}\BibitemShut
  {NoStop}%
\bibitem [{\citenamefont {Stratonovich}(1967)}]{stratonovich1967}%
  \BibitemOpen
  \bibfield  {author} {\bibinfo {author} {\bibfnamefont {R.~L.}\ \bibnamefont
  {Stratonovich}},\ }\href@noop {} {\emph {\bibinfo {title} {Topics in the
  {{Theory}} of {{Random Noise}}}}},\ Vol.~\bibinfo {volume} {2}\ (\bibinfo
  {publisher} {{Gordon and Breach}},\ \bibinfo {address} {New York},\ \bibinfo
  {year} {1967})\BibitemShut {NoStop}%
\bibitem [{\citenamefont {Thomsen}\ and\ \citenamefont
  {Wiseman}(2002)}]{thomsen2002}%
  \BibitemOpen
  \bibfield  {author} {\bibinfo {author} {\bibfnamefont {L.~K.}\ \bibnamefont
  {Thomsen}}\ and\ \bibinfo {author} {\bibfnamefont {H.~M.}\ \bibnamefont
  {Wiseman}},\ }\bibfield  {title} {\bibinfo {title} {Atom laser coherence and
  its control via feedback},\ }\href
  {https://doi.org/10.1103/PhysRevA.65.063607} {\bibfield  {journal} {\bibinfo
  {journal} {Phys. Rev. A}\ }\textbf {\bibinfo {volume} {65}},\ \bibinfo
  {pages} {063607} (\bibinfo {year} {2002})}\BibitemShut {NoStop}%
\bibitem [{\citenamefont {Henry}(1982)}]{henry1982}%
  \BibitemOpen
  \bibfield  {author} {\bibinfo {author} {\bibfnamefont {C.}~\bibnamefont
  {Henry}},\ }\bibfield  {title} {\bibinfo {title} {Theory of the linewidth of
  semiconductor lasers},\ }\href {https://doi.org/10.1109/JQE.1982.1071522}
  {\bibfield  {journal} {\bibinfo  {journal} {IEEE J. Quantum Electron.}\
  }\textbf {\bibinfo {volume} {18}},\ \bibinfo {pages} {259} (\bibinfo {year}
  {1982})}\BibitemShut {NoStop}%
\bibitem [{\citenamefont {Minganti}\ \emph {et~al.}(2021)\citenamefont
  {Minganti}, \citenamefont {Arkhipov}, \citenamefont {Miranowicz},\ and\
  \citenamefont {Nori}}]{minganti2021}%
  \BibitemOpen
  \bibfield  {author} {\bibinfo {author} {\bibfnamefont {F.}~\bibnamefont
  {Minganti}}, \bibinfo {author} {\bibfnamefont {I.~I.}\ \bibnamefont
  {Arkhipov}}, \bibinfo {author} {\bibfnamefont {A.}~\bibnamefont
  {Miranowicz}},\ and\ \bibinfo {author} {\bibfnamefont {F.}~\bibnamefont
  {Nori}},\ }\bibfield  {title} {\bibinfo {title} {Liouvillian spectral
  collapse in the {{Scully-Lamb}} laser model},\ }\href
  {https://doi.org/10.1103/PhysRevResearch.3.043197} {\bibfield  {journal}
  {\bibinfo  {journal} {Phys. Rev. Res.}\ }\textbf {\bibinfo {volume} {3}},\
  \bibinfo {pages} {043197} (\bibinfo {year} {2021})}\BibitemShut {NoStop}%
\bibitem [{\citenamefont {Risken}(1996)}]{risken1996}%
  \BibitemOpen
  \bibfield  {author} {\bibinfo {author} {\bibfnamefont {H.}~\bibnamefont
  {Risken}},\ }\href@noop {} {\emph {\bibinfo {title} {The {{Fokker-Planck}}
  Equation: Methods of Solution and Applications}}},\ \bibinfo {edition}
  {second edition}\ ed.,\ \bibinfo {series} {Springer Series in Synergetics}\
  No.~\bibinfo {number} {18}\ (\bibinfo  {publisher} {Springer},\ \bibinfo
  {address} {Berlin Heidelberg},\ \bibinfo {year} {1996})\BibitemShut {NoStop}%
\bibitem [{\citenamefont {Albert}\ and\ \citenamefont
  {Jiang}(2014)}]{albert2014}%
  \BibitemOpen
  \bibfield  {author} {\bibinfo {author} {\bibfnamefont {V.~V.}\ \bibnamefont
  {Albert}}\ and\ \bibinfo {author} {\bibfnamefont {L.}~\bibnamefont {Jiang}},\
  }\bibfield  {title} {\bibinfo {title} {Symmetries and conserved quantities in
  {{Lindblad}} master equations},\ }\href
  {https://doi.org/10.1103/PhysRevA.89.022118} {\bibfield  {journal} {\bibinfo
  {journal} {Phys. Rev. A}\ }\textbf {\bibinfo {volume} {89}},\ \bibinfo
  {pages} {022118} (\bibinfo {year} {2014})}\BibitemShut {NoStop}%
\bibitem [{\citenamefont {Albert}(2018)}]{albert2018}%
  \BibitemOpen
  \bibfield  {author} {\bibinfo {author} {\bibfnamefont {V.~V.}\ \bibnamefont
  {Albert}},\ }\emph {\bibinfo {title} {Lindbladians with Multiple Steady
  States: Theory and Applications}},\ \href@noop {} {Ph.D. thesis},\ \bibinfo
  {school} {Yale} (\bibinfo {year} {2018})\BibitemShut {NoStop}%
\bibitem [{\citenamefont {Bu{\v c}a}\ \emph {et~al.}(2022)\citenamefont {Bu{\v
  c}a}, \citenamefont {Booker},\ and\ \citenamefont {Jaksch}}]{buca2022}%
  \BibitemOpen
  \bibfield  {author} {\bibinfo {author} {\bibfnamefont {B.}~\bibnamefont
  {Bu{\v c}a}}, \bibinfo {author} {\bibfnamefont {C.}~\bibnamefont {Booker}},\
  and\ \bibinfo {author} {\bibfnamefont {D.}~\bibnamefont {Jaksch}},\
  }\bibfield  {title} {\bibinfo {title} {Algebraic theory of quantum
  synchronization and limit cycles under dissipation},\ }\href
  {https://doi.org/10.21468/scipostphys.12.3.097} {\bibfield  {journal}
  {\bibinfo  {journal} {SciPost Phys.}\ }\textbf {\bibinfo {volume} {12}},\
  \bibinfo {pages} {097} (\bibinfo {year} {2022})}\BibitemShut {NoStop}%
\bibitem [{\citenamefont {{Blume-Kohout}}\ \emph {et~al.}(2010)\citenamefont
  {{Blume-Kohout}}, \citenamefont {Ng}, \citenamefont {Poulin},\ and\
  \citenamefont {Viola}}]{blume-kohout2010}%
  \BibitemOpen
  \bibfield  {author} {\bibinfo {author} {\bibfnamefont {R.}~\bibnamefont
  {{Blume-Kohout}}}, \bibinfo {author} {\bibfnamefont {H.~K.}\ \bibnamefont
  {Ng}}, \bibinfo {author} {\bibfnamefont {D.}~\bibnamefont {Poulin}},\ and\
  \bibinfo {author} {\bibfnamefont {L.}~\bibnamefont {Viola}},\ }\bibfield
  {title} {\bibinfo {title} {Information preserving structures: {{A}} general
  framework for quantum zero-error information},\ }\href
  {https://doi.org/10.1103/PhysRevA.82.062306} {\bibfield  {journal} {\bibinfo
  {journal} {Phys. Rev. A}\ }\textbf {\bibinfo {volume} {82}},\ \bibinfo
  {pages} {062306} (\bibinfo {year} {2010})}\BibitemShut {NoStop}%
\bibitem [{\citenamefont {Lidar}\ \emph {et~al.}(1998)\citenamefont {Lidar},
  \citenamefont {Chuang},\ and\ \citenamefont {Whaley}}]{lidar1998}%
  \BibitemOpen
  \bibfield  {author} {\bibinfo {author} {\bibfnamefont {D.~A.}\ \bibnamefont
  {Lidar}}, \bibinfo {author} {\bibfnamefont {I.~L.}\ \bibnamefont {Chuang}},\
  and\ \bibinfo {author} {\bibfnamefont {K.~B.}\ \bibnamefont {Whaley}},\
  }\bibfield  {title} {\bibinfo {title} {Decoherence {{Free Subspaces}} for
  {{Quantum Computation}}},\ }\href
  {https://doi.org/10.1103/PhysRevLett.81.2594} {\bibfield  {journal} {\bibinfo
   {journal} {Phys. Rev. Lett.}\ }\textbf {\bibinfo {volume} {81}},\ \bibinfo
  {pages} {2594} (\bibinfo {year} {1998})}\BibitemShut {NoStop}%
\bibitem [{\citenamefont {Kempe}\ \emph {et~al.}(2001)\citenamefont {Kempe},
  \citenamefont {Bacon}, \citenamefont {Lidar},\ and\ \citenamefont
  {Whaley}}]{kempe2001}%
  \BibitemOpen
  \bibfield  {author} {\bibinfo {author} {\bibfnamefont {J.}~\bibnamefont
  {Kempe}}, \bibinfo {author} {\bibfnamefont {D.}~\bibnamefont {Bacon}},
  \bibinfo {author} {\bibfnamefont {D.~A.}\ \bibnamefont {Lidar}},\ and\
  \bibinfo {author} {\bibfnamefont {K.~B.}\ \bibnamefont {Whaley}},\ }\bibfield
   {title} {\bibinfo {title} {Theory of decoherence-free fault-tolerant
  universal quantum computation},\ }\href
  {https://doi.org/10.1103/physreva.63.042307} {\bibfield  {journal} {\bibinfo
  {journal} {Phys. Rev. A}\ }\textbf {\bibinfo {volume} {63}},\ \bibinfo
  {pages} {042307} (\bibinfo {year} {2001})}\BibitemShut {NoStop}%
\bibitem [{\citenamefont {Aleiner}\ \emph {et~al.}(2012)\citenamefont
  {Aleiner}, \citenamefont {Altshuler},\ and\ \citenamefont
  {Rubo}}]{aleiner2012a}%
  \BibitemOpen
  \bibfield  {author} {\bibinfo {author} {\bibfnamefont {I.~L.}\ \bibnamefont
  {Aleiner}}, \bibinfo {author} {\bibfnamefont {B.~L.}\ \bibnamefont
  {Altshuler}},\ and\ \bibinfo {author} {\bibfnamefont {Y.~G.}\ \bibnamefont
  {Rubo}},\ }\bibfield  {title} {\bibinfo {title} {Radiative coupling and weak
  lasing of exciton-polariton condensates},\ }\href
  {https://doi.org/10.1103/PhysRevB.85.121301} {\bibfield  {journal} {\bibinfo
  {journal} {Phys. Rev. B}\ }\textbf {\bibinfo {volume} {85}},\ \bibinfo
  {pages} {121301(R)} (\bibinfo {year} {2012})}\BibitemShut {NoStop}%
\bibitem [{\citenamefont {Kalinin}\ and\ \citenamefont
  {Berloff}(2019)}]{kalinin2019}%
  \BibitemOpen
  \bibfield  {author} {\bibinfo {author} {\bibfnamefont {K.~P.}\ \bibnamefont
  {Kalinin}}\ and\ \bibinfo {author} {\bibfnamefont {N.~G.}\ \bibnamefont
  {Berloff}},\ }\bibfield  {title} {\bibinfo {title} {Polaritonic network as a
  paradigm for dynamics of coupled oscillators},\ }\href
  {https://doi.org/10.1103/PhysRevB.100.245306} {\bibfield  {journal} {\bibinfo
   {journal} {Phys. Rev. B}\ }\textbf {\bibinfo {volume} {100}},\ \bibinfo
  {pages} {245306} (\bibinfo {year} {2019})}\BibitemShut {NoStop}%
\bibitem [{\citenamefont {Ji}\ \emph {et~al.}(2023)\citenamefont {Ji},
  \citenamefont {Zhong}, \citenamefont {Ge}, \citenamefont {Beaudoin},
  \citenamefont {Sagnes}, \citenamefont {Raineri}, \citenamefont
  {{El-Ganainy}},\ and\ \citenamefont {Yacomotti}}]{ji2023}%
  \BibitemOpen
  \bibfield  {author} {\bibinfo {author} {\bibfnamefont {K.}~\bibnamefont
  {Ji}}, \bibinfo {author} {\bibfnamefont {Q.}~\bibnamefont {Zhong}}, \bibinfo
  {author} {\bibfnamefont {L.}~\bibnamefont {Ge}}, \bibinfo {author}
  {\bibfnamefont {G.}~\bibnamefont {Beaudoin}}, \bibinfo {author}
  {\bibfnamefont {I.}~\bibnamefont {Sagnes}}, \bibinfo {author} {\bibfnamefont
  {F.}~\bibnamefont {Raineri}}, \bibinfo {author} {\bibfnamefont
  {R.}~\bibnamefont {{El-Ganainy}}},\ and\ \bibinfo {author} {\bibfnamefont
  {A.~M.}\ \bibnamefont {Yacomotti}},\ }\bibfield  {title} {\bibinfo {title}
  {Tracking exceptional points above the lasing threshold},\ }\href
  {https://doi.org/10.1038/s41467-023-43874-z} {\bibfield  {journal} {\bibinfo
  {journal} {Nat. Commun.}\ }\textbf {\bibinfo {volume} {14}},\ \bibinfo
  {pages} {8304} (\bibinfo {year} {2023})}\BibitemShut {NoStop}%
\bibitem [{\citenamefont {Fruchart}\ \emph {et~al.}(2021)\citenamefont
  {Fruchart}, \citenamefont {Hanai}, \citenamefont {Littlewood},\ and\
  \citenamefont {Vitelli}}]{fruchart2021}%
  \BibitemOpen
  \bibfield  {author} {\bibinfo {author} {\bibfnamefont {M.}~\bibnamefont
  {Fruchart}}, \bibinfo {author} {\bibfnamefont {R.}~\bibnamefont {Hanai}},
  \bibinfo {author} {\bibfnamefont {P.~B.}\ \bibnamefont {Littlewood}},\ and\
  \bibinfo {author} {\bibfnamefont {V.}~\bibnamefont {Vitelli}},\ }\bibfield
  {title} {\bibinfo {title} {Non-reciprocal phase transitions},\ }\href
  {https://doi.org/10.1038/s41586-021-03375-9} {\bibfield  {journal} {\bibinfo
  {journal} {Nature}\ }\textbf {\bibinfo {volume} {592}},\ \bibinfo {pages}
  {363} (\bibinfo {year} {2021})}\BibitemShut {NoStop}%
\bibitem [{\citenamefont {Prosen}(2012)}]{prosen2012}%
  \BibitemOpen
  \bibfield  {author} {\bibinfo {author} {\bibfnamefont {T.}~\bibnamefont
  {Prosen}},\ }\bibfield  {title} {\bibinfo {title} {{{PT-symmetric}} quantum
  {{Liouvillian}} dynamics},\ }\href
  {https://doi.org/10.1103/PhysRevLett.109.090404} {\bibfield  {journal}
  {\bibinfo  {journal} {Phys. Rev. Lett.}\ }\textbf {\bibinfo {volume} {109}},\
  \bibinfo {pages} {090404} (\bibinfo {year} {2012})}\BibitemShut {NoStop}%
\bibitem [{\citenamefont {Nakanishi}\ and\ \citenamefont
  {Sasamoto}(2026)}]{nakanishi2026}%
  \BibitemOpen
  \bibfield  {author} {\bibinfo {author} {\bibfnamefont {Y.}~\bibnamefont
  {Nakanishi}}\ and\ \bibinfo {author} {\bibfnamefont {T.}~\bibnamefont
  {Sasamoto}},\ }\bibfield  {title} {\bibinfo {title} {Lindbladian {{PT}} phase
  transitions},\ }\href {https://doi.org/10.1088/1361-6455/ae99b9} {\bibfield
  {journal} {\bibinfo  {journal} {J. Phys. B At. Mol. Opt. Phys.}\ }\textbf
  {\bibinfo {volume} {59}},\ \bibinfo {pages} {172001} (\bibinfo {year}
  {2026})}\BibitemShut {NoStop}%
\bibitem [{\citenamefont {Nakanishi}\ \emph {et~al.}(2026)\citenamefont
  {Nakanishi}, \citenamefont {Hanai},\ and\ \citenamefont
  {Sasamoto}}]{nakanishi2026a}%
  \BibitemOpen
  \bibfield  {author} {\bibinfo {author} {\bibfnamefont {Y.}~\bibnamefont
  {Nakanishi}}, \bibinfo {author} {\bibfnamefont {R.}~\bibnamefont {Hanai}},\
  and\ \bibinfo {author} {\bibfnamefont {T.}~\bibnamefont {Sasamoto}},\
  }\bibfield  {title} {\bibinfo {title} {Continuous {{Time Crystals}} as a {{PT
  Symmetric State}} and the {{Emergence}} of {{Critical Exceptional Points}}},\
  }\href {https://doi.org/10.1103/4w7y-kjzk} {\bibfield  {journal} {\bibinfo
  {journal} {Phys. Rev. Lett.}\ }\textbf {\bibinfo {volume} {136}},\ \bibinfo
  {pages} {250404} (\bibinfo {year} {2026})}\BibitemShut {NoStop}%
\bibitem [{\citenamefont {Thomas}\ and\ \citenamefont
  {Senthilvelan}(2026)}]{thomas2026}%
  \BibitemOpen
  \bibfield  {author} {\bibinfo {author} {\bibfnamefont {N.}~\bibnamefont
  {Thomas}}\ and\ \bibinfo {author} {\bibfnamefont {M.}~\bibnamefont
  {Senthilvelan}},\ }\bibfield  {title} {\bibinfo {title}
  {Normal-mode-splitting-induced synchronization blockade in coupled quantum
  van der {{Pol}} oscillators},\ }\href {https://doi.org/10.1103/xnzk-6rhr}
  {\bibfield  {journal} {\bibinfo  {journal} {Phys. Rev. A}\ }\textbf {\bibinfo
  {volume} {113}},\ \bibinfo {pages} {L020202} (\bibinfo {year}
  {2026})}\BibitemShut {NoStop}%
\bibitem [{\citenamefont {Milburn}\ \emph {et~al.}(2015)\citenamefont
  {Milburn}, \citenamefont {Doppler}, \citenamefont {Holmes}, \citenamefont
  {Portolan}, \citenamefont {Rotter},\ and\ \citenamefont
  {Rabl}}]{milburn2015}%
  \BibitemOpen
  \bibfield  {author} {\bibinfo {author} {\bibfnamefont {T.~J.}\ \bibnamefont
  {Milburn}}, \bibinfo {author} {\bibfnamefont {J.}~\bibnamefont {Doppler}},
  \bibinfo {author} {\bibfnamefont {C.~A.}\ \bibnamefont {Holmes}}, \bibinfo
  {author} {\bibfnamefont {S.}~\bibnamefont {Portolan}}, \bibinfo {author}
  {\bibfnamefont {S.}~\bibnamefont {Rotter}},\ and\ \bibinfo {author}
  {\bibfnamefont {P.}~\bibnamefont {Rabl}},\ }\bibfield  {title} {\bibinfo
  {title} {General description of quasiadiabatic dynamical phenomena near
  exceptional points},\ }\href {https://doi.org/10.1103/PhysRevA.92.052124}
  {\bibfield  {journal} {\bibinfo  {journal} {Phys. Rev. A}\ }\textbf {\bibinfo
  {volume} {92}},\ \bibinfo {pages} {052124} (\bibinfo {year}
  {2015})}\BibitemShut {NoStop}%
\bibitem [{\citenamefont {Schumer}\ \emph {et~al.}(2022)\citenamefont
  {Schumer}, \citenamefont {Liu}, \citenamefont {Leshin}, \citenamefont {Ding},
  \citenamefont {Alahmadi}, \citenamefont {Hassan}, \citenamefont {Nasari},
  \citenamefont {Rotter}, \citenamefont {Christodoulides}, \citenamefont
  {LiKamWa},\ and\ \citenamefont {Khajavikhan}}]{schumer2022}%
  \BibitemOpen
  \bibfield  {author} {\bibinfo {author} {\bibfnamefont {A.}~\bibnamefont
  {Schumer}}, \bibinfo {author} {\bibfnamefont {Y.~G.~N.}\ \bibnamefont {Liu}},
  \bibinfo {author} {\bibfnamefont {J.}~\bibnamefont {Leshin}}, \bibinfo
  {author} {\bibfnamefont {L.}~\bibnamefont {Ding}}, \bibinfo {author}
  {\bibfnamefont {Y.}~\bibnamefont {Alahmadi}}, \bibinfo {author}
  {\bibfnamefont {A.~U.}\ \bibnamefont {Hassan}}, \bibinfo {author}
  {\bibfnamefont {H.}~\bibnamefont {Nasari}}, \bibinfo {author} {\bibfnamefont
  {S.}~\bibnamefont {Rotter}}, \bibinfo {author} {\bibfnamefont {D.~N.}\
  \bibnamefont {Christodoulides}}, \bibinfo {author} {\bibfnamefont
  {P.}~\bibnamefont {LiKamWa}},\ and\ \bibinfo {author} {\bibfnamefont
  {M.}~\bibnamefont {Khajavikhan}},\ }\bibfield  {title} {\bibinfo {title}
  {Topological modes in a laser cavity through exceptional state transfer},\
  }\href {https://doi.org/10.1126/science.abl6571} {\bibfield  {journal}
  {\bibinfo  {journal} {Science}\ }\textbf {\bibinfo {volume} {375}},\ \bibinfo
  {pages} {884} (\bibinfo {year} {2022})}\BibitemShut {NoStop}%
\bibitem [{\citenamefont {Sun}\ and\ \citenamefont {Yi}(2023)}]{sun2023}%
  \BibitemOpen
  \bibfield  {author} {\bibinfo {author} {\bibfnamefont {K.}~\bibnamefont
  {Sun}}\ and\ \bibinfo {author} {\bibfnamefont {W.}~\bibnamefont {Yi}},\
  }\bibfield  {title} {\bibinfo {title} {Chiral state transfer under
  dephasing},\ }\href {https://doi.org/10.1103/PhysRevA.108.013302} {\bibfield
  {journal} {\bibinfo  {journal} {Phys. Rev. A}\ }\textbf {\bibinfo {volume}
  {108}},\ \bibinfo {pages} {013302} (\bibinfo {year} {2023})}\BibitemShut
  {NoStop}%
\bibitem [{\citenamefont {Sun}\ and\ \citenamefont {Yi}(2024)}]{sun2024}%
  \BibitemOpen
  \bibfield  {author} {\bibinfo {author} {\bibfnamefont {K.}~\bibnamefont
  {Sun}}\ and\ \bibinfo {author} {\bibfnamefont {W.}~\bibnamefont {Yi}},\
  }\bibfield  {title} {\bibinfo {title} {Encircling the {{Liouvillian}}
  exceptional points: A brief review},\ }\href
  {https://doi.org/10.1007/s43673-024-00129-3} {\bibfield  {journal} {\bibinfo
  {journal} {AAPPS Bull.}\ }\textbf {\bibinfo {volume} {34}},\ \bibinfo {pages}
  {22} (\bibinfo {year} {2024})}\BibitemShut {NoStop}%
\bibitem [{\citenamefont {Ohadi}\ \emph {et~al.}(2016)\citenamefont {Ohadi},
  \citenamefont {Gregory}, \citenamefont {Freegarde}, \citenamefont {Rubo},
  \citenamefont {Kavokin}, \citenamefont {Berloff},\ and\ \citenamefont
  {Lagoudakis}}]{ohadi2016a}%
  \BibitemOpen
  \bibfield  {author} {\bibinfo {author} {\bibfnamefont {H.}~\bibnamefont
  {Ohadi}}, \bibinfo {author} {\bibfnamefont {R.~L.}\ \bibnamefont {Gregory}},
  \bibinfo {author} {\bibfnamefont {T.}~\bibnamefont {Freegarde}}, \bibinfo
  {author} {\bibfnamefont {Y.~G.}\ \bibnamefont {Rubo}}, \bibinfo {author}
  {\bibfnamefont {A.~V.}\ \bibnamefont {Kavokin}}, \bibinfo {author}
  {\bibfnamefont {N.~G.}\ \bibnamefont {Berloff}},\ and\ \bibinfo {author}
  {\bibfnamefont {P.~G.}\ \bibnamefont {Lagoudakis}},\ }\bibfield  {title}
  {\bibinfo {title} {Nontrivial {{Phase Coupling}} in {{Polariton
  Multiplets}}},\ }\href {https://doi.org/10.1103/PhysRevX.6.031032} {\bibfield
   {journal} {\bibinfo  {journal} {Phys. Rev. X}\ }\textbf {\bibinfo {volume}
  {6}},\ \bibinfo {pages} {031032} (\bibinfo {year} {2016})}\BibitemShut
  {NoStop}%
\bibitem [{\citenamefont {Liang}\ \emph {et~al.}(2026)\citenamefont {Liang},
  \citenamefont {Alnatah}, \citenamefont {Yao}, \citenamefont {Beaumariage},
  \citenamefont {West}, \citenamefont {Baldwin}, \citenamefont {Gupta},
  \citenamefont {Pfeiffer}, \citenamefont {Berloff},\ and\ \citenamefont
  {Snoke}}]{liang2026}%
  \BibitemOpen
  \bibfield  {author} {\bibinfo {author} {\bibfnamefont {S.}~\bibnamefont
  {Liang}}, \bibinfo {author} {\bibfnamefont {H.}~\bibnamefont {Alnatah}},
  \bibinfo {author} {\bibfnamefont {Q.}~\bibnamefont {Yao}}, \bibinfo {author}
  {\bibfnamefont {J.}~\bibnamefont {Beaumariage}}, \bibinfo {author}
  {\bibfnamefont {K.}~\bibnamefont {West}}, \bibinfo {author} {\bibfnamefont
  {K.}~\bibnamefont {Baldwin}}, \bibinfo {author} {\bibfnamefont
  {A.}~\bibnamefont {Gupta}}, \bibinfo {author} {\bibfnamefont {L.~N.}\
  \bibnamefont {Pfeiffer}}, \bibinfo {author} {\bibfnamefont {N.~G.}\
  \bibnamefont {Berloff}},\ and\ \bibinfo {author} {\bibfnamefont {D.~W.}\
  \bibnamefont {Snoke}},\ }\bibfield  {title} {\bibinfo {title}
  {Mirror-mediated long-range coupling and robust phase locking of spatially
  separated exciton-polariton condensates},\ }\bibfield  {journal} {\bibinfo
  {journal} {arXiv:2506.20924}\ }\href
  {https://doi.org/10.48550/arXiv.2506.20924} {10.48550/arXiv.2506.20924}
  (\bibinfo {year} {2026})\BibitemShut {NoStop}%
\bibitem [{\citenamefont {Berloff}\ \emph {et~al.}(2017)\citenamefont
  {Berloff}, \citenamefont {Silva}, \citenamefont {Kalinin}, \citenamefont
  {Askitopoulos}, \citenamefont {T{\"o}pfer}, \citenamefont {Cilibrizzi},
  \citenamefont {Langbein},\ and\ \citenamefont {Lagoudakis}}]{berloff2017a}%
  \BibitemOpen
  \bibfield  {author} {\bibinfo {author} {\bibfnamefont {N.~G.}\ \bibnamefont
  {Berloff}}, \bibinfo {author} {\bibfnamefont {M.}~\bibnamefont {Silva}},
  \bibinfo {author} {\bibfnamefont {K.}~\bibnamefont {Kalinin}}, \bibinfo
  {author} {\bibfnamefont {A.}~\bibnamefont {Askitopoulos}}, \bibinfo {author}
  {\bibfnamefont {J.~D.}\ \bibnamefont {T{\"o}pfer}}, \bibinfo {author}
  {\bibfnamefont {P.}~\bibnamefont {Cilibrizzi}}, \bibinfo {author}
  {\bibfnamefont {W.}~\bibnamefont {Langbein}},\ and\ \bibinfo {author}
  {\bibfnamefont {P.~G.}\ \bibnamefont {Lagoudakis}},\ }\bibfield  {title}
  {\bibinfo {title} {Realizing the classical {{XY Hamiltonian}} in polariton
  simulators},\ }\href {https://doi.org/10.1038/nmat4971} {\bibfield  {journal}
  {\bibinfo  {journal} {Nat. Mater.}\ }\textbf {\bibinfo {volume} {16}},\
  \bibinfo {pages} {1120} (\bibinfo {year} {2017})}\BibitemShut {NoStop}%
\bibitem [{\citenamefont {Cao}\ \emph {et~al.}(2019)\citenamefont {Cao},
  \citenamefont {Chriki}, \citenamefont {Bittner}, \citenamefont {Friesem},\
  and\ \citenamefont {Davidson}}]{cao2019}%
  \BibitemOpen
  \bibfield  {author} {\bibinfo {author} {\bibfnamefont {H.}~\bibnamefont
  {Cao}}, \bibinfo {author} {\bibfnamefont {R.}~\bibnamefont {Chriki}},
  \bibinfo {author} {\bibfnamefont {S.}~\bibnamefont {Bittner}}, \bibinfo
  {author} {\bibfnamefont {A.~A.}\ \bibnamefont {Friesem}},\ and\ \bibinfo
  {author} {\bibfnamefont {N.}~\bibnamefont {Davidson}},\ }\bibfield  {title}
  {\bibinfo {title} {Complex {{Lasers}} with {{Controllable Coherence}}},\
  }\href {https://doi.org/10.1038/s42254-018-0010-6} {\bibfield  {journal}
  {\bibinfo  {journal} {Nat. Rev. Phys.}\ }\textbf {\bibinfo {volume} {1}},\
  \bibinfo {pages} {156} (\bibinfo {year} {2019})}\BibitemShut {NoStop}%
\bibitem [{\citenamefont {Ortega-Taberner}\ \emph {et~al.}()\citenamefont
  {Ortega-Taberner}, \citenamefont {Cotton},\ and\ \citenamefont
  {Eastham}}]{data}%
  \BibitemOpen
  \bibfield  {author} {\bibinfo {author} {\bibfnamefont {C.}~\bibnamefont
  {Ortega-Taberner}}, \bibinfo {author} {\bibfnamefont {R.~E.}\ \bibnamefont
  {Cotton}},\ and\ \bibinfo {author} {\bibfnamefont {P.~R.}\ \bibnamefont
  {Eastham}},\ }\bibfield  {title} {\bibinfo {title} {Supporting code and
  data},\ }\href {https://doi.org/10.5281/zenodo.xxxxxxx}
  {10.5281/zenodo.xxxxxxx}\BibitemShut {NoStop}%
\end{thebibliography}%

\onecolumngrid

\appendix

\section{Semiclassical approximation to the Liouvillian spectrum of a single laser}
\label{appendix:single_laser}
The adjoint Liouvillian of a laser is given by
\begin{align}
    \mathcal{L}^\ddagger(x) =& i[\omega a^\dagger a,x] + A \mathcal{S}^{-1}(\mathcal{D}[a^\dagger]^\ddagger(x))+C\mathcal{D}[a]^\ddagger(x),
\end{align}
where $\mathcal{S}(O) = O + \{ a^\dagger a,O\}B/2A$, and we used that it is equal to its adjoint. The difficulty in treating this Liouvillian analytically resides in the non-linearity of the saturated gain. The inverse of the saturation operator can be expanded in Liouville space can be expanded as
\begin{align}
\mathcal{S} =&\mathbb{I}\otimes \mathbb{I} +\frac{B}{2A}(\hat{n} \otimes \mathbb{I} + \mathbb{I} \otimes \hat{n}) \nonumber \\
=& \sum_{nm} \left(1 + \frac{B}{2A}(n+m) \right) \ket{n}\bra{n} \otimes \ket{m}\bra{m}.
\end{align}
Since it is diagonal, the saturation operator is simply
\begin{align}
\mathcal{S}^{-1} =& \sum_{nm} \frac{1}{1 + (n+m)B/2A }\ket{n}\bra{n} \otimes \ket{m}\bra{m}.
\end{align}
We now consider the action of this operator on $a^{\dagger \,p}a^q$, which will serve as basis to construct the eigenmodes,
\begin{align}
\mathcal{S}^{-1}(a^{\dagger \,p}a^q) =& \sum_{n\ge p,m\ge q} \frac{1}{1 + (n+m)B/2A  }\ket{n}\bra{n} a^{\dagger \,p}a^q \ket{m}\bra{m} \nonumber \\
=& \sum_{n,m=0} \frac{1}{1 + (n+m)B/2A +(p+q)B/2A  }\ket{n+p}\bra{n+p} a^{\dagger \,p}a^q \ket{m+q}\bra{m+q} \nonumber \\
=&  \sum_{n,m=0} \frac{1}{1 + (n+m)B/2A +(p+q)B/2A  }a^{\dagger \,p}\ket{n}\bra{n}  \ket{m}\bra{m} a^q \nonumber \\
=& \sum_{n=0} \frac{1}{1 + nB/A +(p+q)B/2A  }a^{\dagger \,p}\ket{n}\bra{n} a^q.
\end{align}
We then use that
\begin{align}
\ket{n}\bra{n} =& \frac{1}{n!} a^{\dagger \, n}\ket{0}\bra{0} a^n  \nonumber\\
=&\frac{1}{n!} a^{\dagger \, n}\left( \sum_{k=0}^\infty \frac{(-1)^k a^{\dagger\,k} a^k}{k!} \right)a^n ,
\end{align}
where we used the identity $\ket{0}\bra{0} = :e^{-a^\dagger a}:$. We obtain the normal ordered expression
\begin{align}
\mathcal{S}^{-1}(a^{\dagger \,p}a^q) =& \sum_{n=0}^\infty \frac{1}{nB/A+(p+q)B/2A+1  }\frac{1}{n!} \sum_{k=0}^\infty \frac{(-1)^k }{k!} a^{\dagger \, n+p+k} a^{n+q+k} .
\end{align}
We then shift the $k$ index and reverse the sums,
\begin{align}
\mathcal{S}^{-1}(a^{\dagger \,p}a^q) =& \sum_{n=0}^\infty \frac{1}{nB/A+(p+q)B/2A+1  }\frac{1}{n!} \sum_{k=n}^\infty \frac{(-1)^{k-n} }{(k-n)!} a^{\dagger \, p+k} a^{q+k} \nonumber \\
=& \sum_{k=0}^\infty \sum_{n=0}^k \frac{1}{nB/A+(p+q)B/2A     +1  }\mqty(k\\n)\frac{(-1)^{k-n} }{k!} a^{\dagger \, p+k} a^{q+k} .
\end{align}
such that we can perform the $n$ sum. To do so we make use of the partial franction decomposition of the inverse binomial coefficient,
\begin{align}
1/\mqty(z+n\\n) = \sum_{i=1}^n (-1)^{i-1} \mqty(n\\i)\frac{i}{z+i},
\end{align}
to compute the sum

\begin{align}
\sum_{n=0}^k  \mqty(k\\n)(-1)^{k-n}\frac{1}{nB/A+1+(p+q)B/2A} =& \sum_{n=1}^{k+1}  \mqty(k\\n-1)(-1)^{k-n+1}\frac{1}{(n-1)B/A+1+(p+q)B/2A} \nonumber \\
=&\sum_{n=1}^{k+1}  \mqty(k+1\\n)(-1)^{k-n+1}\frac{n}{k+1}\frac{1}{n-1+A/B+(p+q)/2}\frac{A}{B} \nonumber\\
=& \frac{A/B}{k+1}(-1)^{k} \sum_{n=1}^{k+1}  \mqty(k+1\\n)(-1)^{n-1}\frac{n}{n+(A/B-1)+(p+q)/2} \nonumber\\
=& \frac{A/B}{k+1}(-1)^{k} / \mqty((p+q)/2 + k+A/B\\ k+1 ) \nonumber\\
=& (-1)^k\frac{A}{B} \frac{\Gamma(k+2)\Gamma((p+q)/2+A/B)}{(k+1)\Gamma((p+q/2)+k+A/B+1)} \nonumber\\
=&(-1)^k \frac{A}{B}\frac{\Gamma(k+1)\Gamma((p+q)/2+A/B)}{\Gamma((p+q/2)+k+A/B+1)} \nonumber\\
=&(-1)^{k} \frac{A}{B}B(k+1,(p+q)/2+A/B)
\end{align}
where $B(z_1,z_2) = \Gamma(z_1)\Gamma(z_2)/\Gamma(z_1+z_2)$ is the Beta function.  Finally we have the action of the saturation operator,
\begin{align}
\mathcal{S}^{-1}(a^{\dagger \,p}a^q) =& \sum_{k=0}^\infty \frac{(-1)^{k}}{k!}\frac{A}{B} B(k+1,(p+q)/2+A/B)a^{\dagger \, p+k} a^{q+k} .
\end{align}
The dissipators act as
\begin{align}
    \mathcal{D}[a]^\ddagger(a^{\dagger \, p}a^q) =& \frac{1}{2}a^\dagger[a^{\dagger p}a^q,a]-\frac{1}{2}[a^{\dagger p}a^q,a^\dagger]a, \nonumber\\
    =& -\frac{1}{2} (p+q) a^{\dagger p}a^q,
\end{align}
and,
\begin{align}
    \mathcal{D}[a^\dagger]^\ddagger(a^{\dagger \, p}a^q) =& \frac{1}{2}a[a^{\dagger p}a^q,a^\dagger]-\frac{1}{2}[a^{\dagger p}a^q,a]a^\dagger,\nonumber \\
    =& \frac{1}{2}q a a^{\dagger \, p}a^{q-1}+\frac{1}{2} p a^{\dagger \, p-1}a^q a^\dagger \nonumber\\
    =& pq a^{\dagger \, p-1}a^{q-1} +\frac{p+q}{2}a^{\dagger p}a^q ,
\end{align}
and the Hamiltonian term,
\begin{align}
i[\omega a^\dagger a, a^{\dagger \, p}a^q] =& i\omega \left( a^\dagger  [a, a^{\dagger \, p}a^q] +[a^\dagger , a^{\dagger \, p}a^q]a\right)\nonumber\\
=& i\omega (p-q)a^{\dagger \, p}a^q.
\end{align}
Putting everything together we obtain the Liouvillian
\begin{align}
\mathcal{L}^\ddagger(a^{\dagger \, p}a^q) =& i\omega(p-q)a^{\dagger \,p}a^q + A\mathcal{S}^{-1}\left( pq a^{\dagger \, p-1}a^{q-1} +\frac{p+q}{2}a^{\dagger p}a^q \right)  -\frac{1}{2}C (p+q) a^{\dagger p}a^q \nonumber\\
=& \frac{A^2}{B} pq \sum_{k=0}^\infty \frac{(-1)^{k}}{k!} B(k+1,(p+q)/2+A/B-1)a^{\dagger \, p-1+k} a^{q-1+k} \nonumber\\
&+ \frac{A^2}{B} \frac{p+q}{2} \sum_{k=0}^\infty \frac{(-1)^{k}}{k!} B(k+1,(p+q)/2+A/B)a^{\dagger \, p+k} a^{q+k}\nonumber  \\
& -\frac{1}{2}C (p+q)a^{\dagger p}a^q.
\end{align}
Here we can explicitely see the effect of the $U(1)$ symmetry discussed in the main text such that the Liouvillian only couples modes $a^{\dagger \, p}a^q$ with a constant $p-q = \nu$.

In order to approximate the eigenvalues and eigenmodes of the Liouvillian it is easier to work with the phase space representation,
\begin{align}
\bra{z}\mathcal{L}^\ddagger(a^{\dagger \, p}a^q) \ket{z} 
=& \frac{A^2}{B}  pq \sum_{k=0}^\infty (-1)^{k} \frac{\Gamma((p+q)/2+A/B-1)}{\Gamma((p+q)/2+k+A/B)} \abs{z}^{2k} z^{*\, p-1}z^{q-1}\nonumber\\
&+ \frac{A^2}{B} \frac{p+q}{2} \sum_{k=0}^\infty (-1)^{k} \frac{\Gamma((p+q)/2+A/B)}{\Gamma((p+q)/2+k+1+A/B)} \abs{z}^{2k} z^{*\, p}z^{q}\nonumber\\
& -\frac{1}{2}C (p+q)z^{*\, p}z^{q}\nonumber \\
=& 2\frac{A^2}{B} \frac{pq}{p+q+2(A/B-1)} z^{*\, p-1}z^{q-1} {}_1F_1(1;(p+q)/2+A/B;-\abs{z}^2)\nonumber\\
&+ \frac{A^2}{B}\frac{p+q}{p+q+2A/B} z^{*\, p}z^{q} {}_1F_1(1;(p+q)/2+1+A/B;-\abs{z}^2)\nonumber\\
& -\frac{1}{2}C (p+q)z^{*\, p}z^{q} ,
\end{align}
where we used the definition of the confluent hypergeometric function ${}_1F_1(a;b;x)$. In the following we use $\mu = (p+q)/2, \nu = p-q$ to simplify notation, and set $A=B$, as that is the case explored in the main text. We then obtain   
\begin{align}
\bra{z}\mathcal{L}^\ddagger(a^{\dagger \, \mu+k/2}a^{\mu-\nu/2}) \ket{z} 
=& A \frac{\mu^2-\nu^2/4}{\mu} \abs{z}^{2\mu-2} e^{-i\nu\theta} {}_1F_1(1;\mu+1;-\abs{z}^2)\nonumber\\
&+ A\frac{\mu}{\mu+1} \abs{z}^{2\mu} e^{-i\nu\theta} {}_1F_1(1;\mu+2;-\abs{z}^2)\nonumber\\
& -\mu C\abs{z}^{2\mu} e^{-i\nu\theta} ,
\end{align}
where we used that $z = \abs{z}e^{i\theta}$. Using the identity 
\begin{align}
{}_1F_q(1;\mu+1;-\abs{z}^2) = 1 -\frac{\abs{z}^2}{\mu+1}{}_1F_q(1;\mu+2;-\abs{z}^2),
\end{align}
we can simplify the result to
\begin{align}
\bra{z}\mathcal{L}^\ddagger(a^{\dagger \, \mu+k/2}a^{\mu-\nu/2}) \ket{z} 
=& A \frac{\mu^2-\nu^2/4}{\mu} \abs{z}^{2\mu-2} e^{-i\nu\theta} \nonumber\\
&-A \frac{\mu^2-\nu^2/4}{\mu} \abs{z}^{2\mu-2} e^{-i\nu\theta}\frac{\abs{z}^2}{\mu+1}{}_1F_1(1;\mu+2;-\abs{z}^2)\nonumber\\
&+ A\frac{\mu}{\mu+1} \abs{z}^{2\mu} e^{-i\nu\theta} {}_1F_1(1;\mu+2;-\abs{z}^2)\nonumber\\
& -\mu C\abs{z}^{2\mu} e^{-i\nu\theta}\nonumber \\
=&  \abs{z}^{2\mu}e^{-i\nu\theta} \left(A \frac{\mu^2-\nu^2/4}{\mu}\frac{1}{\abs{z}^2}-\mu C \right) \nonumber\\
&+ A \abs{z}^{2\mu}e^{-i\nu\theta} \frac{\nu^2/4}{(\mu+1)\mu}{}_1F_1(1;\mu+2;-\abs{z}^2)
\end{align}
As discussed above different $\nu$ corresponds to different symmetry blocks and therefore the eigenmodes of the Liouvillian are of the form $\psi_{\nu,\lambda} = \sum_{\mu}c^\nu_{\lambda,\mu}a^{\dagger \, \mu+\nu/2}a^{\mu-\nu/2}$, such that
\begin{align}
\mathcal{L}^\ddagger(\psi_{\nu,\lambda}) = \lambda \psi_{\nu,\lambda}.
\end{align}
The eigenvalue equation results in
\begin{align}
\lambda \sum_{\mu}c_{\lambda,\mu} \abs{z}^{2\mu}e^{-i\nu \theta} =& \sum_{\mu}c_{\lambda,\mu}  \abs{z}^{2\mu}e^{-i\nu\theta} \left(A \frac{\mu^2-\nu^2/4}{\mu}\frac{1}{\abs{z}^2}-\mu C \right) \nonumber\\
&+ \sum_{\mu}c_{\lambda,\mu} A \abs{z}^{2\mu}e^{-i\nu\theta} \frac{\nu^2/4}{(\mu+1)\mu}{}_1F_1(1;\mu+2;-\abs{z}^2).
\end{align}
So far the eigenvalue equation is exact. However, in order to find solutions we need to approximate the hypergeometric function. Since the relevant phase space in the classical limit is that for $\abs{z} \gg 1$, we can use the asymptotic expansion of the hypergeometric function,
\begin{align}
{}_1F_1(1;\mu+2;-r^2) =& \sum_{n=0}^\infty (-1)^n \frac{\Gamma(\mu+2)}{\Gamma(\mu+1-n)}\abs{z}^{-2-2n} \nonumber\\
\approx& \left(\mu+1 \right) \abs{z}^{-2}-\left(\mu+1\right)\mu \abs{z}^{-4} + O(\abs{z}^{-6})
\end{align}
such that
\begin{align}
\lambda \sum_{\mu}c_{\lambda,\mu} \abs{z}^{2\mu}\approx& \sum_{\mu}c_{\lambda,\mu}  \abs{z}^{2\mu} \left(A \frac{\mu^2-\nu^2/4}{\mu}\frac{1}{\abs{z}^2}-\mu C \right) \nonumber\\
&+ \sum_{\mu}c_{\lambda,\mu} A \abs{z}^{2\mu}\frac{\nu^2/4}{\mu}\left(\frac{1}{\abs{z}^2} - \mu \frac{1}{\abs{z}^4} \right)\nonumber \\
=&\sum_{\mu}c_{\lambda,\mu}  \abs{z}^{2\mu} \mu \left(A \frac{1}{\abs{z}^2}- C \right) -\sum_{\mu}c_{\lambda,\mu} A \abs{z}^{2\mu}\frac{\nu^2}{4} \frac{1}{\abs{z}^4}  
\end{align}
Expressing the eigenmode as $\psi_{\nu,\lambda} = R_{\nu,\lambda}(r)e^{-i\nu\theta}$, with $r = \abs{z}^2$, we can express the above eigenvalue equation as a differential equation 
\begin{align}
\lambda R_{\nu,\lambda}(r) = A \partial_r R_{\nu,\lambda}(r) - C r\partial_r R_{\nu,\lambda}(r) -A \frac{\nu^2}{4}\frac{1}{r^2}R_{\nu,\lambda}(r),
\end{align}
with solutions
\begin{align}
R_{\nu,\lambda}(r) = e^{-\nu^2/4r} r^{C\nu^2/4A}(A-Cr)^{-\lambda/C - C\nu^2/4A}.
\end{align}
In order for the solutions to be physical, the exponent of $(A-Cr)$ should be a positive integer, as otherwise the solution can present singularities or branch points. This leads to Eq.~\eqref{eq:anespec}, 
which is the behavior observed numerically, and matches what has been previously obtained with other semiclassical approaches. Note that while this is an asymptotic approximation, higher order terms are proportional to $\nu^2$, and therefore this result is exact for the $\nu=0$ case. For $k=1$, in particular, the eigenmode corresponds to $\psi_{0,-C} \propto a^{\dagger}a - A/C$, which corresponds to the deviation of the amplitude with respect to the classical result $n_0 = A/C$. The eigenmodes of the limit cycle branch, for $k=0$, are
\begin{align}
R_{\nu,\lambda}(r) = e^{-\nu^2/4r} r^{C\nu^2/4A}.
\end{align}
For large $r$ and $A/C$, i.e. for the phase space near the classical solution, both powers are suppressed and $R_{\nu,\nu\omega} (r) \approx  1$. Meaning that the full eigenstates in this limit correspond simply to multiples of the phase operator $\psi_{\nu,\nu\omega} \approx e^{-i\nu \theta}$, as shown in the numerics. 

\section{Symmetry of the Husimi Q function}\label{appendix:sym}

In the single laser case the Liouvillian commutes with the $U(1)$ symmetry
\begin{align}
    \mathcal{U} = e^{-i\theta \hat{n}} \otimes e^{i\theta \hat{n}},
\end{align}
which acts on the eigenstates of the Liouvillian as
\begin{align}
    \mathcal{U}(\psi_\nu) = e^{i\nu\theta}\psi_\nu.
\end{align}
The Husimi Q function is defined as
\begin{align}
    Q(\alpha) = \bra{\alpha} \psi_\nu \ket{\alpha}.
\end{align}
Using the symmetry,
\begin{align}
    Q(\alpha) =& \bra{\alpha} \mathcal{U}(\psi_\nu) \ket{\alpha}e^{-i\nu \theta} \nonumber\\
    =& \bra{\alpha e^{i\phi}} \psi_\nu \ket{\alpha e^{i\theta}}e^{-i\nu\theta} \nonumber\\
    =& Q(\alpha e^{i\theta})e^{-i\theta},
\end{align}
where we used $e^{i\theta \hat{n}}\ket{\alpha} = \ket{\alpha e^{i\theta }}$. Choosing $\theta = -\arg{\alpha}$ we obtain the property
\begin{align}
    Q(\alpha) = Q(\abs{\alpha}) e^{i\nu\arg(\alpha)}.
\end{align}

Similarly, in the case with two coupled lasers, the Liouvillian commutes with the collective $U(1)$ symmetry
\begin{align}
    \mathcal{U} = e^{-i\theta ( \hat{n}_1+\hat{n}_2)} \otimes e^{i\theta (\hat{n}_1+\hat{n}_2)},
\end{align}
which acts on the eigenstates of the Liouvillian as
\begin{align}
    \mathcal{U}(\psi_\nu) = e^{i\nu\theta}\psi_\nu.
\end{align}
Defining the collective modes
\begin{align}
    & b_1 = \cos(\theta) a_1 + \sin(\theta) e^{i\phi} a_2 \nonumber\\
    & b_2 = \sin(\theta)e^{-i\phi} a_1 - \cos(\theta) a_2.
\end{align}
the two-body Husimi Q function is
\begin{align}
    Q(\beta_1,\beta_2) = \bra{\beta_1,\beta_2} \psi_\nu \ket{\beta_1,\beta_2},
\end{align}
and the symmetry can also be expressed as $\mathcal{U} = e^{-i\theta ( \hat{m}_1+\hat{m}_2)} \otimes e^{i\theta (\hat{m}_1+\hat{m}_2)}$, where $\hat{m}_j = b_j^\dagger b_j$. Using now the symmetry,
\begin{align}
    Q(\beta_1,\beta_2) =& \bra{\beta_1,\beta_2} \mathcal{U}(\psi_\nu) \ket{\beta_1,\beta_2}e^{-i\nu \theta} \nonumber\\
    =& \bra{\beta_1 e^{i\phi},\beta_2 e^{i\phi}} \psi_\nu \ket{\beta_1 e^{i\phi},\beta_2 e^{i\phi}}e^{-i\nu\theta} \nonumber\\
    =& Q(\beta_1 e^{i\theta},\beta_2 e^{i\theta})e^{-i\theta},
\end{align}
where we again used $e^{i\theta \hat{m}}\ket{\beta} = \ket{\beta e^{i\theta} }$. Choosing $\theta = -\arg{\beta_2}$ we obtain 
\begin{align}
    Q(\beta_1,\beta_2) = Q(\beta_1,\abs{\beta_2}) e^{i\nu\arg(\beta_2)},
\end{align}
which we exploit in the main text to visualize the Husimi Q in three dimensions.

\end{document}